\documentclass[twoside]{book}
\usepackage{quotchap}

\usepackage{minitoc}

\usepackage{fancyhdr}
\usepackage{amssymb}
\usepackage{amsmath}
\usepackage{bm}
\usepackage{graphicx}
\usepackage{fancybox}
\usepackage{mathdots}
\usepackage{rotating}
\usepackage{makeidx}
\usepackage{color}
\usepackage{framed} 
\usepackage{wasysym}
\usepackage{gravity} 

\usepackage{graphicx,epsf,epsfig,amssymb} 
\usepackage{cite} 
\usepackage{hyperref}

\def\nn{\nonumber}

\def\be{\begin{equation}}
\def\ee{\end{equation}}
\def\beq{\begin{eqnarray}}
\def\eeq{\end{eqnarray}}

\def\ii{{\rm i}}
\def\nn{\nonumber}
\def\f{\frac}

\def\ii{{\rm i}}
\def\nn{\nonumber}

\newcommand{\beqn}{\begin{eqnarray*}}
\newcommand{\eeqn}{\end{eqnarray*}}
\newcommand{\p}{\partial}

\def\op{ \ $ }
\def\cl{$ \ }

\def\l2{\op \bf L^2\cl}

\def\l{{\cal L}}

\makeindex
\begin{document}

\dominitoc 
\setcounter{minitocdepth}{1} 

\pagenumbering{roman}
\thispagestyle{empty}
%

\vspace*{.5in} 

\begin{center} 
\Huge \sc
A Black-Hole Primer:\\ 
\huge \sc 
Particles, Waves,\\
\huge \sc 
Critical Phenomena and Superradiant Instabilities

\vspace*{1in} 
\huge \sc 
Bad Honnef School ``GR$@$99''
\vspace*{1in} 

\huge \sc 
Emanuele Berti
 \\

\vspace*{.2in} 

\normalsize \rm 
(14 October 2014)

\end{center}

%
\tableofcontents 

\pagenumbering{arabic}
%
\chapter{Introduction}
\label{ch:intro} 
\allowdisplaybreaks
\minitoc

These informal notes were prepared for a lecture on black-hole physics
delivered at the DPG Physics School ``General Relativity $@$ 99'',
organized by Gerhard Sch\"afer and Clifford M.~Will and held on Sep
14--19 2014 at the Physikzentrum in Bad Honnef, Germany. The goal of
the notes is to introduce some key ideas in black hole physics that
highlight an intimate connection between the dynamics of particles and
waves in black-hole spacetimes. More specifically, I will show that
the geodesic motion of {\em particles} is related to the
characteristic properties of {\em wave} scattering in black hole
backgrounds. I will introduce the proper oscillation modes of a black
hole (``quasinormal modes'') and illustrate how they are related with
the stability of circular orbits for massless particles in the
geometrical optics limit. Finally, I will explain the basic ideas
behind superradiance, and I will give a concise introduction to ``hot
topics'' in current research. The prerequisite background is a
knowledge of physics at the advanced undergraduate level, namely
nonrelativistic quantum mechanics and General Relativity at the level
of Hartle \cite{Hartle}, Schutz \cite{Schutz} or Carroll
\cite{Carroll}. In particular, I will assume some familiarity with the
Einstein field equations, the Schwarzschild metric and the Kerr
metric.

I will focus on the core physics, rather than the mathematics. So --
whenever given the choice -- I will consider the simplest prototype
problem (one that can quickly be solved with pen and paper), providing
references to generalizations that involve heavier calculations but no
substantial new physics.

In Chapter \ref{ch:particles}, after a short review of the derivation
of the geodesic equations from a variational principle, I will focus
on geodesics in spherically symmetric spacetimes (more in particular,
the Schwarzschild spacetime). I will discuss orbital stability in
terms of Lyapunov exponents, then I will apply the formalism to
compute the principal Lyapunov exponent for circular null geodesics in
Schwarzschild.

In Chapter \ref{ch:waves} I will introduce a simple prototype for
black-hole perturbations induced by a ``test'' field, i.e. a massive
scalar field. I will separate variables for the corresponding wave
equation (the Klein-Gordon equation), and then I will discuss two
techniques to solve the associated eigenvalue problem: Leaver's
continued fraction method and the WKB approximation. Last but not
least, I will show how the application of the same techniques to
rotating (Kerr) black holes leads to the interesting possibility of
superradiant amplification.

In Chapter \ref{ch:applications} I will use this ``theoretical
minimum'' to introduce some exciting ideas in black hole physics that
have been a focus of recent research, in particular: (1) critical
phenomena in black-hole binary encounters, and (2) the idea that
astrophysical measurements of black-hole spins can constrain the mass
of light bosonic fields (via superradiant instabilities).

For further reading on black holes I recommend Shapiro and
Teukolsky's {\em Black holes, white dwarfs, and neutron stars}
\cite{Shapiro:1983du}, Frolov and Novikov's {\em Black hole physics:
  Basic concepts and new developments} \cite{Frolov:1998wf},
Chandrasekhar's {\em The Mathematical Theory of Black Holes}
\cite{MTB} and a recent overview article by Teukolsky
\cite{Teukolsky:2014vca}.  For reviews of black-hole perturbation
theory, see \cite{Kokkotas:1999bd,Nollert:1999ji,Berti:2009kk}.
Throughout these notes, unless otherwise stated, I will use
geometrical units ($G=c=1$) and I will follow the sign conventions of
Misner, Thorne and Wheeler \cite{Misner:1974qy}.

\section{Newtonian Black Holes?}
\label{sec:newtBHs} 

Popular articles on black holes often mention that a Newtonian analog
of the black-hole concept (a ``dark star'') was first proposed by
Michell in 1783.  Michell's idea was simple: the escape velocity of an
object at the surface $R$ of a star (or planet) with mass $M$ is
\be
v^2_{\rm esc}=\frac{2GM}{R}\,.
\ee
If we consider light as a corpuscle traveling at speed $c$, light can
not escape to infinity whenever $v_{\rm esc}>c$. Therefore the
condition for existence of ``dark stars'' in Newtonian mechanics is
\be
\frac{2GM}{c^2R}\geq 1\,. \label{BHformation}
\ee

Remarkably, a similar situation holds in Einstein theory. The field
equations tell us that there exist vacuum, spherically symmetric
solutions of the field equations describing a source of mass $M$ such
that the (areal) radius of the horizon -- the region from which even
light can not escape -- corresponds to the equal sign in
(\ref{BHformation}). They also tell us that the redshift of light
emitted from the horizon is infinite: the object is perceived as
completely black. A major difference is that in Newtonian gravity
particles can cross the surface $r=2GM/c^2$, while in General
Relativity this surface marks a causal boundary in spacetime.

What popular books usually do not address is the following question:
can Michell's solutions exist in nature? In General Relativity, the
most compact stars are made out of incompressible matter and they
correspond to the Buchdahl limit $R/M=9/4$ (see e.g.~\cite{Schutz}),
but they are not black holes! In Newtonian mechanics, a naive argument
tells us that as we pile up more and more material of constant density
$\rho_0$, the ratio $M/R$ increases:
\be
\frac{M}{R}=\frac{4}{3}\pi R^2 \rho_0\,.
\ee
This equation would seem to suggest that Michell's dark stars {\em
  could} indeed form. Unfortunately, we forgot to include the binding
energy $U$:
\be
U=\int -\frac{GM\, dM}{r}=-\int \f{G}{r} \left(\frac{4}{3}\pi\,r^3
\rho_0\right) 4\pi r^2\rho_0 dr=-\frac{16G\pi^2}{15}\rho_0^2R^5\,.
\ee
Once we do, the total mass $M_T$ of the hypothetical dark star is
given by the rest mass $M$ plus the binding energy:
\be
\frac{M_T}{R}=\frac{4}{3}\pi R^2 \rho_0-\frac{16 G\pi^2}{15c^2}\rho_0^2R^4
=\frac{M}{R}\left[1-\frac{3}{5}\frac{G}{c^2}\frac{M}{R}\right]
\leq \frac{5c^2}{12G}\,,
\ee
where the upper limit is obtained by maximizing the function in the
range (\ref{BHformation}). Thus, the ``dark star criterion''
(\ref{BHformation}) is never satisfied, even for the unrealistic case
of constant-density matter. In fact, the endpoint of Newtonian
gravitational collapse depends very sensitively on the equation of
state, even in spherical symmetry
\cite{Glass:1980,1980AnPhy.130...99M}.

Despite the physical impossibility to construct Newtonian black holes
with ``ordinary'' matter, the genesis of the idea is historically
interesting. The email from Steve Detweiler to James Fry reported in
Box~\ref{box:Detweiler} (reproduced here with permission from Steve
Detweiler - thanks!) has some amusing remarks on Reverend Michell's
contributions to science.

\index{Michell!Detweiler|(}
\begin{BOX}{Steve Detweiler's email to James Fry on Reverend Michell's ``black holes'' (and other amusing things)}
\label{box:Detweiler}

Date: Tue, 25 Aug 98 08:11:25 -0400 \\
From: Steve Detweiler \\
To: James Fry \\
Subject: Re: Newtonian black holes \\

Jim,

The Reverend John Michell in a letter to his good friend Henry  
Cavendish that was published in Transactions of the Royal Society in  
1784.  Here is an excerpt of a summary I made in a letter to a  
friend--I think it's amusing!

The reference is "On the Means of discovering the Distance,  
Magnitude, etc. of the Fixed Stars, in consequence of the Diminution  
of the Velocity of their Light, in case such a Diminution should be  
found to take place in any of them, and such other Data should be  
procured from Observations, as would be farther necessary for that  
Purpose", by the Rev. John Michell, B.D.F.R.S., in a letter to Henry  
Cavendish, Esq. F.R.S. and A.S., Philosophical Transactions of the  
Royal Society of London, Vol. LXXIV, 35-57 (1784).  We sure don't  
title papers the way they used to.

I was at Yale when I found out about this article, so I called up  
their rare books library and was informed that 1784 really wasn't so  
long ago and that the volume I wanted was on the stacks in the main  
library.  I just checked it out, they had to glue in a new due date  
sticker in back, and I had it on my desk for six months.

Michell's idea is to determine the distance to the stars by  
measuring the speed of light from the stars--the further away the  
star is, the slower the light would be moving.  He conceded that  
there would need to be improvement in prism quality for this to be  
possible, but he thought it was within reach.  It's an interesting  
paper, the arguments are all geometrical, where there should be an  
equation he writes it out long hand, all the s's look like f's, and  
it's fun to read.  Somewhere in the middle of the paper (paragraph  
16), he notes that if a star were of the same density as the sun but  
with a radius 500 times larger, then the light would be pulled back  
to star and not escape.  He even surmises that we might be able to  
detect such an object if it were in a binary system and we observed  
the motion of the companion.  So we should have celebrated the Black  
Hole Bicentennial back in 1984!

And Michell was an interesting fellow.  He wrote 4 papers that I  
know about, and I think these are in chronological order.  In one  
paper (1760), he notes a correspondence between volcanos and fault  
lines in the earth's crust and first suggested that earthquakes  
might be caused by masses of rocks shifting beneath the earth's  
surface.  In another, Michell (1767) estimates that if the stars  
were just scattered randomly on the celestial sphere then the  
probability that the Pleiades would be grouped together in the sky  
is 1 part in 500,000.  So he concluded that the Pleiades must be a  
stellar system in its own right---this is the first known  
application of statistics to astronomy.  In a third paper, he  
devised a torsion balance and used it to measure the $1/r^2$  
dependence of the force between magnetic poles.  His fourth paper  
has the black hole reference.  After that work, he started a project  
to modify his magnetic-force torsion balance to measure the force of  
gravity between two masses in the laboratory.  Unfortunately, he  
died in the middle of the project. But his good friend, Henry  
Cavendish, stepped in, made important improvements and completed the  
experiment!  In his own paper, Cavendish gives a footnote reference  
to Michell.
\end{BOX}
\index{Michell!Detweiler|)}

\section{The Schwarzschild and Kerr Metrics}
\label{sec:SchwKerr} 

I refer to the textbooks by Hartle \cite{Hartle}, Schutz
\cite{Schutz} or Carroll \cite{Carroll} for introductions to the
Schwarzschild and Kerr spacetimes that provide the necessary
background for these lecture notes. Here I collect the relevant
metrics and a few useful results, mainly to establish notation.
The Schwarzschild metric reads
\be
ds^2=-\left(1-\f{2M}{r}\right)dt^2
+\left(1-\f{2M}{r}\right)^{-1}dr^2
+r^2(d\theta^2+\sin^2\theta d\phi^2)\,,
\ee
where $M$ is the black-hole mass. The Kerr metric for a black-hole
with mass $M$ and angular momentum $J$ reads:
\beq
ds^2&=&-\left(1-\f{2Mr}{\Sigma}\right)dt^2
-\f{4aMr\sin^2\theta}{\Sigma}dtd\phi \\
&+&\f{\Sigma}{\Delta}dr^2
+\Sigma d\theta^2
+\left(r^2+a^2+\f{2Mra^2\sin^2\theta}{\Sigma}\right)
\sin^2\theta d\phi^2\,. \nn
\eeq
where
\be
a\equiv \f{J}{M}
\,,\quad
\Delta\equiv r^2-2Mr+a^2
\,,\quad
\Sigma\equiv r^2+a^2\cos^2\theta\,.
\ee
The event horizon is located at the larger root of $\Delta=0$, namely
\be
r_+=M+\sqrt{M^2-a^2}\,.
\ee
Stationary observers at fixed values of $r$ and $\theta$ rotate with
constant angular velocity
\be\label{Omega}
\Omega=\f{d\phi}{dt}=\f{u^\phi}{u^t}\,.
\ee
For timelike observers, the condition $u_\mu u^\mu=-1$ yields
\be
-1=u_t^2 \left[g_{tt}+2\Omega g_{t\phi}+\Omega^2 g_{\phi\phi}\right]\,,
\ee
where I have used (\ref{Omega}) to eliminate $u^\phi$. The quantity
in square brackets must be negative; but since $g_{\phi\phi}$ is
positive, this is true only if $\Omega_-<\Omega<\Omega_+$, where
$\Omega_{\pm}$ denotes the two roots of the quadratic, i.e.
\be
\Omega_{\pm}=\f{-g_{t\phi}\pm\sqrt{g_{t\phi}^2-g_{tt}g_{\phi\phi}}}{g_{\phi\phi}}\,.
\ee
Note that $\Omega_-=0$ when $g_{tt}=0$,
i.e. $r^2-2Mr+a^2\cos^2\theta=0$, which occurs at
\be
r_0=M+(M^2-a^2\cos^2\theta)^{1/2}\,.
\ee
Observers between $r_0$ and $r_+$ must have $\Omega>0$, i.e. {\em no
  static observers exist within $r_+<r<r_0$}. This is the reason why
the surface $r=r_0$ is called the ``static limit''. The surface $r_0$
is also called the ``boundary of the ergosphere'', because this is the
region within which energy can be extracted from the black hole via
the Penrose process (a close relative of the superradiant
amplification that I will discuss in Chapter \ref{ch:waves}).





%
\chapter{Particles}
\label{ch:particles}
\allowdisplaybreaks
\minitoc

\section{Geodesic Equations from a Variational Principle \label{sec:geodesy}}

In Einstein's theory, particles do not fall under the action of a
force. Their motion corresponds to the trajectory in space-time that
extremizes\footnote{The ``principle of least action'' is a misnomer,
  because the action is not always minimized along a geodesic: it can
  either have a minimum or a saddle point (see \cite{PW} for an
  introduction and \cite{PoissonGray} for a detailed discussion in
  General Relativity, as well as references to the corresponding
  problem in classical mechanics).}  the proper interval between two
events.
Mathematically, this means that the equations of motion (or ``geodesic
equations'') can be found by a variational principle, i.e. by
extremizing the action
\be
\int d\lambda\, 2\l\equiv
\int d\lambda \left(g_{\mu\nu}\frac{dx^{\mu}}{d\lambda}\frac{dx^{\nu}}{d\lambda}\right)\,,
\ee
where the factor of $2$ in the definition of the Lagrangian density
$\l$ was inserted for later convenience. Therefore our task is to
write down the usual Euler-Lagrange equations
\be
\frac{d}{d\lambda}\frac{\partial \l}{\partial \dot{x}^{\alpha}}=\frac{\partial \l}{\partial x^{\alpha}}\,
\ee
for the Lagrangian
\be\label{Lagr}
\l =
\frac{1}{2}g_{\mu\nu}\frac{dx^{\mu}}{d\lambda}\frac{dx^{\nu}}{d\lambda}=
\frac{1}{2}g_{\mu\nu}\dot{x}^{\mu}\dot{x}^{\nu}\,.
\ee
Here I introduced the notation
\be
\dot{x}^{\mu}\equiv \frac{dx^{\mu}}{d\lambda}=p^\mu
\ee
for the particle's four-momentum, a choice that allows us to consider
both massless and massive particles at the same time
(cf.~\cite{Shapiro:1983du,PW} for more details). The ``affine
parameter'' $\lambda$ can be identified with proper time (more
precisely, with $\tau/m_0$) along the timelike geodesic of a particle
of rest mass $m_0$. The canonical momenta associated with this
Lagrangian are
\be
p_\mu=\f{\p \l}{\p \dot x^{\mu}}\,,
\ee
and they satisfy
\be
p_\mu=g_{\mu\nu}\dot x^\nu=g_{\mu\nu}p^\nu\,.
\ee

More explicitly, introducing the usual notation
$\partial_\mu=\frac{\partial}{\partial x^{\mu}}$, one has
\beq
&&\frac{\partial \l}{\partial \dot{x}^{\alpha}}=g_{\mu\alpha}\dot{x}^{\mu}\,,\\
&&\frac{d}{d\lambda}\frac{\partial \l}{\partial \dot{x}^{\alpha}}=g_{\mu\alpha}\ddot{x}^{\mu}+\dot{x}^{\mu} \frac{dg_{\mu\alpha}}{d\lambda}=g_{\mu\alpha}\ddot{x}^{\mu}+\dot{x}^{\mu}\dot{x}^{\nu}\partial_\nu g_{\mu\alpha}\,,\\
&&\frac{\partial \l}{\partial x^{\alpha}}=\frac{1}{2}\partial_\alpha g_{\mu\nu}\dot{x}^{\mu}\dot{x}^{\nu}\,,
\eeq
and therefore
\be
g_{\mu\alpha}\ddot{x}^{\mu}+\dot{x}^{\mu}\dot{x}^{\nu}\partial_\nu g_{\mu\alpha}-\frac{1}{2}\dot{x}^{\mu}\dot{x}^{\nu}\partial_\alpha g_{\mu\nu}=0\,.
\ee
Now, $\mu$ and $\nu$ are dummy indices. By symmetrizing one gets
$2\dot{x}^{\mu}\dot{x}^{\nu}\partial_\nu
g_{\mu\alpha}=\dot{x}^{\mu}\dot{x}^{\nu} (\partial_\nu
g_{\mu\alpha}+\partial_\mu g_{\nu\alpha})$, and finally
\be
2g_{\mu\alpha}\ddot{x}^{\mu}+
\dot{x}^{\mu}\dot{x}^{\nu}
\left(
 \partial_\nu g_{\mu\alpha}
+\partial_\mu g_{\nu\alpha}
-\partial_\alpha g_{\mu\nu}
\right)=0\,.
\ee
From the definition of the Christoffel symbols
\be
\Gamma^\mu_{\,\,\beta \gamma}=
\frac{1}{2}g^{\mu\rho}
\left(
 \partial_\gamma g_{\beta\rho}
+\partial_\beta g_{\gamma\rho}
-\partial_\rho g_{\beta\gamma}
\right)
\label{Eequs}
\ee
it follows that
\be
\ddot{x}^{\mu}+\Gamma^{\mu}_{\,\,\beta\gamma}\dot{x}^{\beta}\dot{x}^{\gamma}=0\,,
\ee
i.e. the usual geodesic equation.

\section{Geodesics in Static, Spherically Symmetric Spacetimes}

In most of these notes I will consider a general static, spherically
symmetric metric (not necessarily the Schwarzschild metric) of the
form
\be 
ds^2=-f(r) dt^2+\frac{1}{h(r)}dr^2+r^2(d\theta ^2+\sin^2\theta d\phi ^2)\,.
\label{Sch}
\ee
This metric can describe (for example) a spherical star, and it
reduces to the Schwarzschild metric (see
e.g.~\cite{Hartle,Schutz,Carroll}) when $f(r)=h(r)=1-2M/r$.
For brevity, in the following I will omit the $r$-dependence of
$f(r)$ and $h(r)$. For this metric, the Lagrangian for geodesic
motion reduces to
\be\label{Lsphersymm}
2\l=g_{\mu\nu}\dot x^\mu\dot x^\nu=
-f\dot{t}^2
+h^{-1}\dot{r}^2
+r^2\left(\dot{\theta}^2+\sin^2\theta\dot{\phi}^2\right)\,.
\ee
The momenta associated with this Lagrangian are
\beq
p_t&=&\f{\p \l}{\p \dot t}=-f\dot t\,,\\
p_r&=&\f{\p \l}{\p \dot r}=h^{-1}\dot r\,, \label{rad1}\\
p_\theta&=&\f{\p \l}{\p \dot \theta}=r^2\dot \theta\,,\\
p_\phi&=&\f{\p \l}{\p \dot \phi}=r^2\sin^2\theta\dot \phi\,.
\eeq
The Hamiltonian is related to the Lagrangian by a Legendre transform:
\be
{\cal H}=
p_\mu \dot x^\mu-\l=
g_{\mu\nu}\dot x^\mu\dot x^\nu
-\l=
\l
\ee
and it is equal to the Lagrangian, as could be anticipated because the
Lagrangian (\ref{Lagr}) is ``purely kinetic'' (there is no potential
energy contribution). Therefore ${\cal H}=\l$ and they are both
constant.

As a consequence of the static, spherically symmetric nature of the
metric, the Lagrangian (\ref{Lsphersymm}) does not depend on $t$ and
$\phi$. Therefore the $t$- and $\phi$-components yield two conserved
quantities $E$ and $L$:
\beq
&&\f{dp_t}{d\lambda}=
\f{d}{d\lambda}\left(\f{\p \l}{\p \dot t}\right)=
\f{\p \l}{\p t}=0
\Longrightarrow
-p_t=f\dot t=E\,,\\
&&\f{dp_\phi}{d\lambda}=
\f{d}{d\lambda}\left(\f{\p \l}{\p \dot \phi}\right)=\f{\p \l}{\p \phi}=0
\Longrightarrow
p_\phi=r^2\sin^2\theta\dot \phi=L\,.
\eeq
The radial equation reads
\beq\label{radial_geods}
&&\f{d p_r}{d\lambda} = 
\f{d}{d\lambda}\left(\f{\p \l}{\p\dot r}\right)=
\f{\p \l}{\p r}\,.
\eeq
The remaining equation tells us that
\be
\f{dp_\theta}{d\lambda}=
\f{d}{d\lambda}(r^2\dot \theta)=
\f{\p \l}{\p \theta}=
r^2\sin\theta\cos\theta \dot\phi^2\,,
\ee
so if we choose $\theta=\pi/2$ when $\dot\theta=0$ it follows that
$\ddot \theta=0$, and the orbit will be confined to the equatorial
plane $\theta=\pi/2$ at all times. In conclusion:
\beq
&f\dot{t}=E\,, \label{tdot}\\
&r^2\dot{\phi}=L\,, \label{phidot}
\eeq
where $E$ is the particle's energy and $L$ is the orbital angular
momentum (in the case of massive particles, these quantities should
actually be interpreted as the particle energy and orbital angular
momentum {\em per unit rest mass}: $E=E_{\rm particle}/m_0$ and
$L=L_{\rm particle}/m_0$, see \cite{Shapiro:1983du} for a clear
discussion). The constancy of the Lagrangian now implies
\be
\f{E^2}{f}-\f{\dot r^2}{h}-\f{L^2}{r^2}=-2{\cal L}=\delta_1\,,
\ee
where the constant $\delta_1=0$ for null orbits, and its value can be
set to be $\delta_1=1$ for timelike orbits by a redefinition of the
affine parameter (i.e., a rescaling of ``proper time units'') along
the geodesic. Therefore the conservation of the Lagrangian is
equivalent to the normalization condition for the particle's
four-velocity: $p_\mu p^\mu=-\delta_1$, where $\delta_1=0$ for massless
particles and $\delta_1=1$ massive particles, respectively.  This is
the familiar ``effective potential'' equation used to study the
qualitative features of geodesic motion:
\beq
&&\dot r^2=V_r\,,\label{rdot}\\
&&V_r\equiv h\left(\f{E^2}{f}-\f{L^2}{r^2}-\delta_1\right)\,,\label{def:Vr}
\eeq
where $\delta_1=0$ ($1$) for null (timelike) geodesics.

\section{Schwarzschild Black Holes}

In the special case of Schwarzschild black holes, $f(r)=h(r)=1-2M/r$,
and the radial equation reduces to
\beq
\dot{r}^2=V_r^{\rm Schw}\equiv E^2-f\left(\frac{L^2}{r^2}+\delta_1\right)\,.
\label{Veff}
\eeq
Carroll \cite{Carroll} rewrites this equation in a form meant to
stress the Newtonian limit:
\be
\f{1}{2}\dot r^2={\cal E}-{\cal V}(r)\,,
\ee
where ${\cal E}\equiv E^2/2$ and
\be
{\cal V}_{\delta_1}(r)=\delta_1\left(\f{1}{2}-\f{M}{r}\right)+\f{L^2}{2r^2}-\f{ML^2}{r^3}\,.
\ee
A more conventional choice (cf. Shapiro and Teukolsky
\cite{Shapiro:1983du} or Misner-Thorne-Wheeler \cite{Misner:1974qy})
is to write
\be
\dot r^2=E^2-2{\cal V}_{\delta_1}(r)
\ee
and to define
\beq
\label{effpot}
&&V_{\rm part}(r)=2{\cal V}_{\delta_1=1}(r)=\left(1-\f{2M}{r}\right)\left(1+\f{L^2}{r^2}\right)\,,\\
\label{effpotphot}
&&V_{\rm phot}(r)=\f{2{\cal V}_{\delta_1=0}(r)}{L^2}=\f{1}{r^2}\left(1-\f{2M}{r}\right)\,.
\eeq
The motion of photons depends only on the impact parameter $b=L/E$ (as
it should, by virtue of the equivalence principle: the same result can
be obtained by a simple rescaling of the affine parameter $\lambda\to
L\lambda$ \cite{Shapiro:1983du}):
\be
\left(\f{dr}{d\lambda}\right)^2=\f{1}{b^2}-V_{\rm phot}(r)\,.
\ee

\begin{figure*}[t]
\begin{center}
\epsfig{file=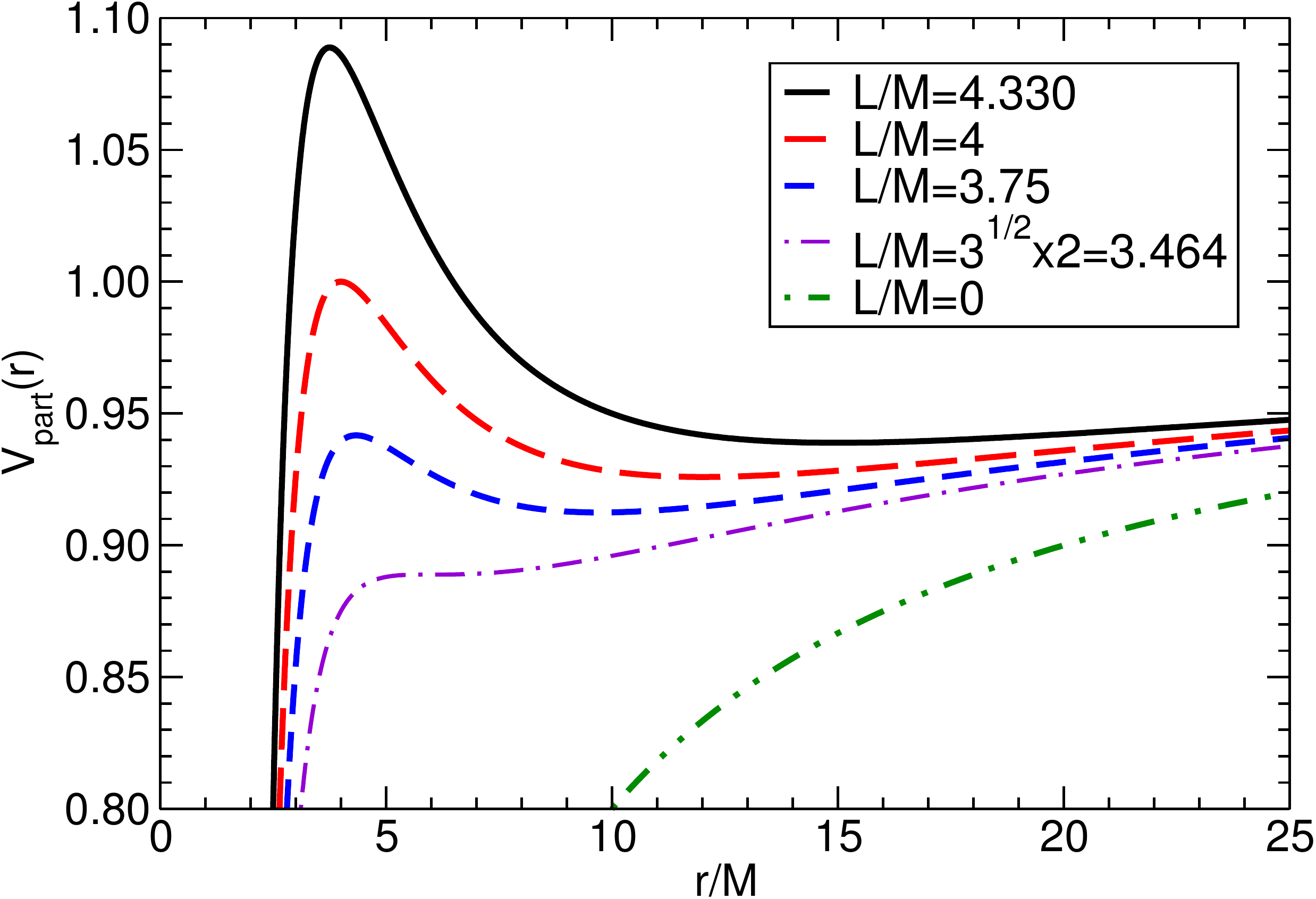,width=\textwidth,angle=0,clip=true}\\
\caption{Effective potential (\ref{effpot}) for massive particles.
  \label{fig:effpot}}
\end{center}
\end{figure*}

\begin{figure*}[t]
\begin{center}
\epsfig{file=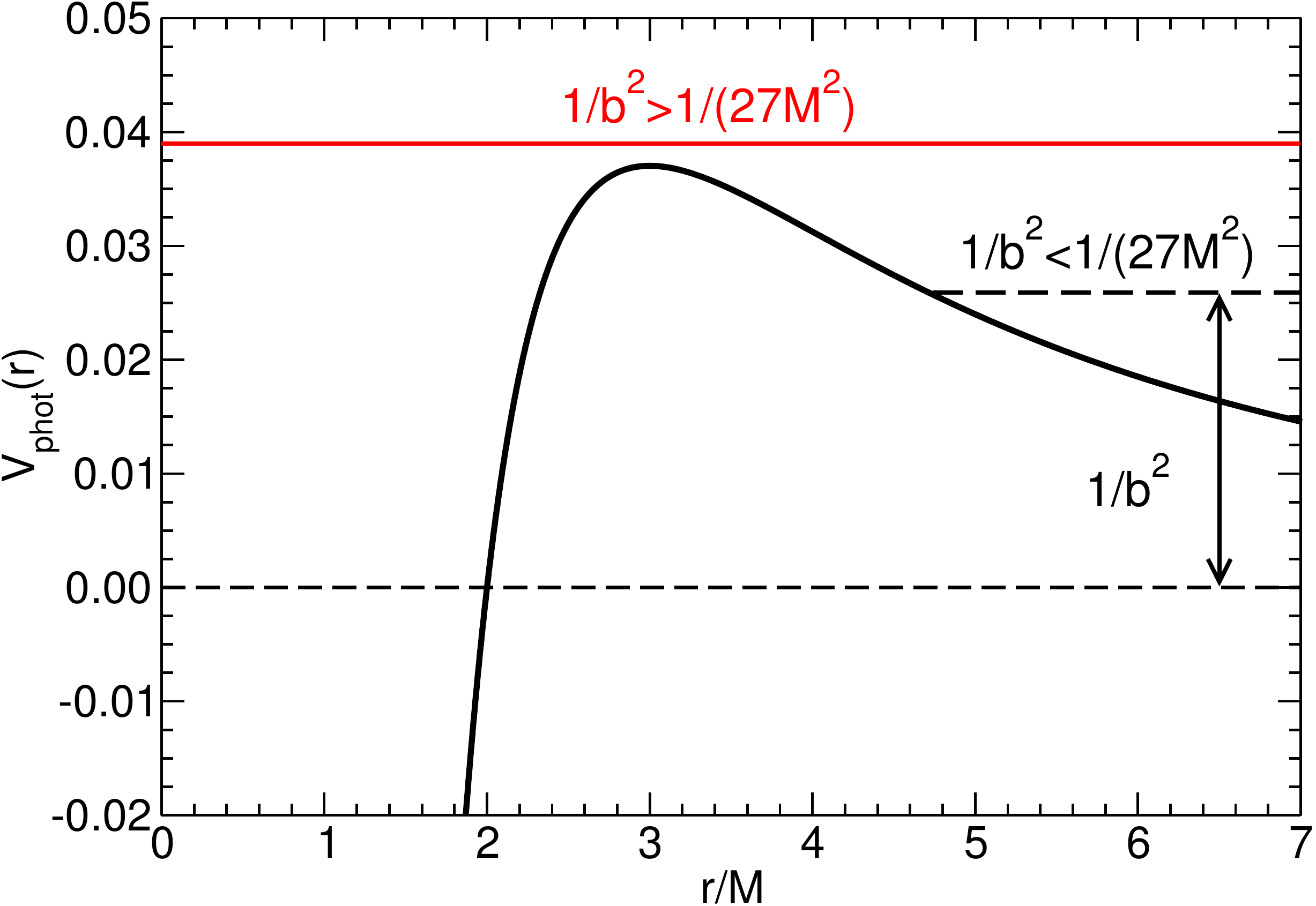,width=\textwidth,angle=0,clip=true}\\
\caption{Effective potential (\ref{effpotphot}) for photons. The
  critical impact parameter (\ref{lightringb}), corresponding to the
  light ring, separates plunging orbits from scattering orbits.
  \label{fig:effpotphot}}
\end{center}
\end{figure*}

The potentials (\ref{effpot}) and (\ref{effpotphot}) are plotted in
Figs.~\ref{fig:effpot} and \ref{fig:effpotphot}, respectively. The
qualitative features of particle motion in these potentials are
treated in most books on General Relativity (see
e.g.~\cite{Misner:1974qy,Hartle,Schutz,Carroll,Shapiro:1983du}), so
there is no need to repeat the discussion here, except for some
important remarks in the next Section that will be useful below.

Note that, according to the definitions above, the derivatives of the
radial particle potential $V_r$ and the derivatives of the potentials
$({\cal V}_{\delta_1},\,V_{\rm part},\,V_{\rm phot})$ have opposite
signs. Therefore unstable orbits (in particular the ``light ring'',
i.e. the unstable photon orbit at $r=3M$, for which $V_{\rm
  phot}''<0$) have $V_r''>0$.
%

\subsection{Circular Geodesics in the Schwarzschild Metric}
By definition, circular geodesics have constant areal radius
coordinate $r$: $\dot{r}=0$ and $\ddot{r}=0$, with the second
condition ensuring that circular orbits remain circular.  From
Eq.~(\ref{rdot}) it follows that $V_r=V_r'=0$.  The condition $(V^{\rm
  Schw}_r)'=0$ is equivalent to
\be
L^2(r-3M)=\delta_1Mr^2\,.\label{Lcircular}
\ee
From this equality one can already draw two conclusions:

\begin{itemize}
\item[(i)] Circular geodesics exist only for $r\geq 3M$.

\item[(ii)] Null circular geodesics ($\delta_1=0$) exist only at
$r=3M$.
\end{itemize}

The angular frequency of the circular orbit (as measured at infinity) is
\be\label{Omegac}
\Omega_c \equiv \frac{d\phi}{dt}=\frac{\dot{\phi}}{\dot{t}}=\frac{fL}{r^2E}\,.
\ee
Eq.~(\ref{Lcircular}) implies $L^2=\delta_1 Mr^2/(r-3M)$, and the
condition $V_r=0$ translates to $E^2=\delta_1rf^2/(r-3M)$. Therefore
we get the relativistic analog of Kepler's law:
\be
\Omega_c^2=\frac{f^2L^2}{r^4E^2}=\frac{M}{r^3}\,.
\ee

To understand the stability of these geodesics under small
perturbations we must look at $(V^{\rm Schw}_r)''$.  After substituting $L^2$ from
Eq.~(\ref{Lcircular}), we get
\be\label{VSpp}
(V^{\rm Schw}_r)''=-\frac{2\delta_1M(r-6M)}{r^3(r-3M)}\,.
\ee
Thus, 

\begin{itemize}
\item[(iii)] Circular geodesics with $r\geq 6M$ are stable (this is
  why $r=6M$ is referred to as the ``innermost stable circular
  orbit'', or ISCO).

\item[(iv)] Circular geodesics with $3M\leq r<6M$ are all unstable. 
\end{itemize}

\subsection{The Critical Impact Parameter}
The effective potential for geodesic motion was given in Eq.~(\ref{Veff}).
Let us consider two extreme cases: $E=1$ and the ultrarelativistic
limit $E\rightarrow \infty$ (where the second case corresponds to
massless, ultrarelativistic particles, and therefore it is equivalent
to considering $\delta_1=0$).

For $E=1$, i.e., particles dropped from rest at infinity, we have
\beq
\dot{r}^2&=&1-f\left(\frac{L^2}{r^2}+1\right)=\frac{2M^3}{r^3}\left(\frac{r^2}{M^2}+\frac{L^2}{M^2}-\frac{1}{2}\frac{L^2}{M^2}\frac{r}{M}\right)\\
&=&\frac{2M^3}{r^3}\left(\frac{r}{M}-\frac{L}{M}\frac{L/M+\sqrt{(L/M)^2-16}}{4}\right)\left(\frac{r}{M}-\frac{L}{M}\frac{L/M-\sqrt{(L/M)^2-16}}{4}\right)\,.\nonumber
\eeq
From this factorization, it is clear that there is a turning point
(i.e., a real value of $r$ for which $\dot{r}=0$) if and only if
$(L/M)^2-16>0$. Furthermore, all turning points lie outside the
horizon (that is located at $r=r_+=2M$). It follows that the critical
angular momentum for capture is
\be\label{Lcrit4M}
L_{\rm crit}=4M\,.
\ee
%

When we consider the capture of massless particles ($\delta_1=0$),
defining an impact parameter $b=L/E$, we get
\be
\dot{r}^2=E^2-f\left(\frac{L^2}{r^2}\right)=\frac{E^2}{r^3}\left(r^3-rfb^2\right)\,.
\ee
To find the turning points we must now solve a cubic
equation. Following Chandrasekhar \cite{MTB}, we look for the critical
$b$ such that the cubic polynomial $r^3-rfb^2=r^3-rb^2+2Mb^2$ has no
real positive roots. Now, it is well known that the roots
$(r_1,\,r_2,\,r_3)$ of the polynomial $ar^3+br^2+cr+d$ satisfy
$r_1+r_2+r_3=-b/a$ and $r_1r_2r_3=-d/a$. We know for sure that there
is always one negative root, because the polynomial tends to $-\infty$
as $r\rightarrow -\infty$ and it is positive (equal to $2Mb^2$) at
$r=0$. Therefore we have two possibilities: (1) one real, negative
root and two complex-conjugate roots, or (2) one real, negative root
and two distinct, real roots. The critical case is when the two real
roots of case (2) degenerate into a single real root. Clearly, at this
point one must have
\be
(r^3-rfb^2)'=3r^2-b^2=0 \quad {\rm or} \quad r=b/\sqrt{3}\,.
\ee
Substituting back into the polynomial, this will only be a root if
\beq\label{lightringb}
b_{\rm crit}&=&3\sqrt{3}M\,,\\
r_{\rm crit}&=&3M\,.
\eeq

The main conclusions of this discussion can be summarized as follows: 

{\bf
\begin{itemize}
\item[(1)] Circular null geodesics (the trajectories of massless
  scalars, photons or gravitons circling the black hole) are located
  at $r=3M$, and they are unstable;

\item[(2)] The critical impact parameter for capture of light rays (or
  gravitons, or massless scalars) corresponds precisely to these
  circular null geodesics.
\end{itemize}
}

As we will see in Chapter \ref{ch:waves}, circular null geodesics are
linked (in the geometric-optics limit) to the oscillation modes of
black holes. To make this understanding quantitative, we need a
measure of the instability rate of circular null geodesics. This
measure relies on the classical theory of orbital stability based on
Lyapunov exponents, that I review below.

\section{Order and Chaos in Geodesic Motion}
\label{sec:lyapunov}

In the theory of dynamical systems, Lyapunov exponents are a measure
of the average rate at which nearby trajectories converge or diverge
in phase space. A positive Lyapunov exponent indicates a divergence
between nearby trajectories, i.e., a high sensitivity to initial
conditions. To investigate geodesic stability in terms of Lyapunov
exponents, we begin with the equations of motion schematically written
as
\be 
\frac{d X_{i}}{dt} = H_{i}(X_{j}) \,. 
\ee
We linearize these equations about a certain orbit $X_i(t)$ and we get
\be 
\frac{d\,  \delta \! X_{i}(t)}{dt} = K_{ij}(t)\,  \delta \! X_{j}(t) \,, 
\ee
where
\be 
K_{ij}(t) = \left. \frac{\partial H_{i}}{\partial X_{j}}\right|_{X_{i}(t)} 
\ee
is the linear stability matrix \cite{Cornish:2003ig}. The solution to the
linearized equations can be written as
\be \label{eq:Lij} 
\delta \! X_{i}(t) = L_{ij}(t)\, \delta\! X_{j}(0) \, 
\ee
in terms of the evolution matrix $L_{ij}(t)$, which must obey
\be \label{evo} \dot L_{ij}(t)=K_{im}(t) L_{mj}(t)\, \ee
as well as $L_{ij}(0) = \delta_{ij}$, so that Eq.~(\ref{eq:Lij}) is
satisfied at $t=0$. The eigenvalues $\lambda_i$ of $L_{ij}$ are called
``Lyapunov exponents''. More precisely, the principal Lyapunov
exponent is the largest of these eigenvalues, such that
\be\label{L} 
\lambda_0 = \lim_{t \rightarrow \infty} \frac{1}{t} \log \left( \frac{ L_{jj} (t)}{L_{jj} (0)}
\right) \, . \ee
%

\subsection{Lyapunov Exponents for Circular Orbits in Static, Spherically Symmetric Spacetimes}

Let us now restrict attention to a class of problems for which one has
a two dimensional phase space of the form $X_i(t)=(p_r, r)$. Let us
focus on circular orbits in static, spherically symmetric
metrics\footnote{Our treatment is actually general enough to study
  circular orbits in stationary spherically symmetric spacetimes, and
  also equatorial circular orbits in stationary spacetimes (including
  the higher-dimensional generalization of rotating black holes, known
  as the Myers-Perry metric): cf.~\cite{Cardoso:2008bp} for
  generalizations to these cases.}.
and linearize the equations of motion with $X_i(t)=(p_r, r)$ about
orbits of constant areal radius $r$. As pointed out by Cornish and
Levin \cite{Cornish:2003ig}, Lyapunov exponents are in general
coordinate-dependent. To measure the orbital instability rate through
the principal Lyapunov exponent $\lambda_0$, a sensible choice is to
use Schwarzschild time $t$, i.e. the time measured by an observer at
infinity. The relevant equations of motion are Eqs.~(\ref{rad1}) and
(\ref{radial_geods}). Linearizing them about circular orbits of radius
$r_c$ yields
\beq
&&\dot{\delta r}=
\left[h'p_r\right]_{r=r_c, p_r=0} \delta r
+h\delta p_r = 
h\delta p_r =
\f{1}{g_{rr}}\delta p_r\,,\\
&&\dot{\delta p_r}=\f{d}{dr}
\left(
\f{\p \l}{\p r}
\right)\delta r\,,
\eeq
where dots denote derivatives with respect to $\lambda$, and it is
understood that the quantities in the linearized equations must be
evaluated at $r=r_c$ and $p_r=0$.  Converting from the affine
parameter $\lambda$ to Schwarzschild time (which implies dividing
everything through by a factor $\dot t$) we get the linear stability
matrix
\be 
K_{ij} = 
\begin{pmatrix}
0 & K_1\\
K_2 & 0
\end{pmatrix}
 \,, 
\ee
where
\beq
K_1&=&\dot{t}^{-1} \frac{d}{dr}\left (\frac{\p \l}{\p r}\right )
\,,\\
K_2&=&\left(\dot{t}\,g_{rr}\right)^{-1}\,.
\eeq
%
%
Therefore, for circular orbits, the principal Lyapunov exponents can be
expressed as
\be
\lambda_0=\pm \sqrt{K_1\,K_2}\,.
\label{pL}\\
\ee
The fact that eigenvalues come in pairs is a reflection of energy
conservation: the volume in phase space is also conserved
\cite{Cornish:2003ig}.  From now on I will drop the $\pm$ sign, and
simply refer to $\lambda_0$ as the ``Lyapunov exponent''. 

From the equations of motion it follows that
\be
\frac{\p \l}{\p r}=\frac{d}{d\lambda}\frac{\p \l}{\p \dot{r}}=\frac{d}{d\lambda}\left(g_{rr} \dot{r}\right )=\dot{r}\frac{d}{dr}(g_{rr}\dot{r})\,,
\ee
where in the second equality I used Eq.~(\ref{Lsphersymm}).  
The definition of $V_r$, Eq.~(\ref{rdot}), implies
\be
V_r'=\f{d}{dr}\left(\dot r^2\right)=2\dot r \f{d}{dr}\dot r
\ee
and 
\be
\dot{r}\frac{d}{dr}(g_{rr}\dot{r})=
g_{rr}\dot r \f{d}{dr}\dot r+\dot r^2 g'_{rr}=
g_{rr}\f{1}{2}V_r'+V_r g'_{rr}=
\f{1}{2g_{rr}}\left[g_{rr}^2 V_r\right]'\,,
\ee
so that finally
\be\label{dLdrgeneral}
\frac{\p \l}{\p r}=\frac{1}{2g_{rr}}\frac{d}{dr}\left ( g_{rr}^2\,V_r\right)\,.
\ee
For circular geodesics $V_r=V_r'=0$ \cite{Bardeen:1972fi}, so Eq.~(\ref{pL})
reduces to
\be
\lambda_0=
\sqrt{ \f{1}{\dot t^2 g_{rr}}\frac{d}{dr}\left(\frac{\p \l}{\p r}\right) } =
\sqrt{ \f{1}{\dot t^2 g_{rr}}\frac{d}{dr}\left[
\frac{1}{2g_{rr}}\frac{d}{dr}\left( g_{rr}^2\,V_r\right) \right] } =
\sqrt{ \f{1}{\dot t^2 g_{rr}} \left(\frac{g_{rr}}{2} V_r'' \right) }
\ee
and finally
\be\label{LyapunovGeneral}
\lambda_0= 
\sqrt{\frac{V_r''}{2 \dot{t}^2}}\,.
\ee
Recall from our discussion of geodesics that unstable orbits have
$V_r''>0$, and correspond to real values of the Lyapunov exponent
$\lambda_0$.  Following Pretorius and Khurana \cite{Pretorius:2007jn},
we can introfuce a critical exponent
\be
\gamma \equiv \frac{\Omega_c}{2\pi \lambda_0} = \frac{T_{\lambda_0}}{T_\Omega}\,,
\ee
where $\Omega_c$ was given in Eq.~(\ref{Omegac}), and in addition we
have introduced a typical orbital timescale $T_\Omega\equiv
2\pi/\Omega_c$ and an instability timescale $T_{\lambda_0}\equiv
1/\lambda_0$ (note that in Ref.~\cite{Cornish:2003ig} the authors use
a different definition of the orbital timescale, $T_{\Omega}\equiv
2\pi/\dot{\phi}$). Then the critical exponent can be written as
\be\label{Gamma}
\gamma= 
\frac{1}{2\pi}
\sqrt{\frac{2 \dot \phi^2}{V_r''}}\,.
\ee
For circular null geodesics in many spacetimes of interest $V_r''>0$
[cf. Eq.~(\ref{VSpp})], which implies instability. 

A quantitative characterization of the instability requires a
calculation of the associated timescale $\lambda_0$. Let us turn to
this calculation for timelike and circular orbits.

\subsubsection{Timelike Geodesics}
From Eq.~(\ref{def:Vr}), the requirement $V_r=V'_r=0$ for circular
orbits yields
\be E^2= \frac{2f^2}{2f-r\,f'}\,,\quad
L^2=\frac{r^3\,f'}{2f-rf'}\,,\label{angmomentumcircular2} \ee
where here and below all quantities are evaluated at the radius of a circular
timelike orbit. Since the energy must be real, we require
\be
2f-rf'\,>\,0\,.\label{timelineconstraint}
\ee
The second derivative of the potential is
\be
V''_r=2\frac{h}{f}\frac{-3ff'/r+2(f')^2-ff''}{2f-rf'}\,,
\label{curvpotentialtimelike}
\ee
and the orbital angular velocity is given by
\be 
\Omega=
\frac{\dot{\phi}}{\dot{t}}=
\sqrt{\frac{f'}{2r}}\,. 
\ee
Using Eqs.~(\ref{LyapunovGeneral}), (\ref{tdot}) and
(\ref{curvpotentialtimelike})
to evaluate the Lyapunov exponent at the circular timelike geodesics,
we get
%
%
\beq\label{lamcirc2}
\lambda_0&=&\frac{1}{\sqrt{2}}\sqrt{-\frac{h}{f}
\left[\frac{3ff'}{r}-2(f')^2+ff''\right]} \, \nonumber\\
&=& \frac{1}{2}\sqrt{\left (2f-rf'\right )V''_r(r})\,.
\eeq
Bearing in mind that $2f-rf'>0$ and that unstable orbits are defined
by $V''_r>0$, we can see that $\lambda_0$ will be real whenever the
orbit is unstable, as expected.

\subsubsection{Null Geodesics}

Circular null geodesics satisfy the conditions:
\beq 
&&V_r=0 \quad \Longrightarrow \quad \frac{E}{L}=\pm \sqrt{\frac{f_c}{r_c^2}}\,,
\label{LElight}\\
&&V_r'= h \left(-\f{E^2}{f^2}f'+\f{2L^2}{r^3}\right) = 
h \f{L^2}{r^3}
\left(-\f{f'r}{f}+2\right)=0\,,
\eeq
and therefore
\be
2f_c=r_cf'_c\,.
\label{cirgeo} 
\ee
Here and below a subscript $c$ means that the quantity in question is
evaluated at the radius $r=r_c$ of a circular null geodesic. An
inspection of (\ref{cirgeo}) shows that circular null geodesics can be
seen as the innermost circular timelike geodesics. By taking another
derivative and using Eq.~(\ref{cirgeo}) we find
\be
V_r''(r_c)=
\f{h_c}{f_c}\frac{L^2}{r_c^4}
\left [2f_c-r_c^2f''_c\right]\,,
\label{V2p}
\ee
and the coordinate angular velocity is
\be
\Omega_c=
\frac{\dot{\phi}}{\dot{t}}=
\left(\frac{f'_c}{2r_c}\right)^{1/2}=
\frac{f_c^{1/2}}{r_c}
\label{Omegacircular}\,.
\ee
Let us define the ``tortoise'' coordinate
\be
\frac{dr}{dr_*}=(hf)^{1/2}\label{tortoise}\,.
\ee
For circular orbits that satisfy Eq.~(\ref{cirgeo}), a simple
calculation shows that
\be
\f{d^2}{dr_*^2}\left(\f{f}{r^2}\right)=-\f{hf}{r^4}\left(2f-r^2f''\right)\,.
\ee
Combining this result with Eq.~(\ref{V2p}), we finally have
\be
\lambda_0= 
-\frac{1}{\sqrt{2}}\sqrt{\frac{r_c^2f_c}{L^2}V_r''(r_c)}\,
= \frac{1}{\sqrt{2}}\sqrt{-\frac{r_c^2}{f_c}\left(\frac{d^2}{dr_*^2}\frac{f}{r^2}\right)_{r=r_c}}\,.\label{lyaponov}
\ee
This is the main result of this Chapter. In the next Chapter I will
show that the rate of divergence of circular null geodesics at the
light ring, as measured by the principal Lyapunov exponent, is equal
(in the geometrical optics limit) to the damping time of black-hole
perturbations induced by any massless bosonic field.

%
\chapter{Waves}
\label{ch:waves}
\allowdisplaybreaks
\minitoc

\section{Massive Scalar Fields in a Spherically Symmetric Spacetime}

Consider now the perturbations induced by a probe {\em field} (rather
than a particle) propagating in a background spacetime, that for the
purpose of this section could be either a black hole or a star. In the
usual perturbative approach we assume that the scalar field $\Phi$
contributes very little to the energy density (i.e. we drop quadratic
terms in the field throughout, assuming that the energy-momentum
tensor of the scalar field is negligible). Then the problem reduces to
that of a scalar field evolving in a fixed background, and the
``true'' metric can be replaced by a solution to the Einstein
equations, that we will assume for simplicity to be spherically
symmetric. Let us therefore start from the metric (\ref{Sch}), which
reduces to the Schwarzschild metric when $f=h=1-2M/r$. The
Klein-Gordon equation in this background reads
\be
\left(\square-\mu^2\right)\Phi = 0\,.\label{kg}
\ee
Here the mass $\mu$ has dimensions of an inverse length, $m=\mu \hbar$
(with $G=c=1$). The necessary conversions can be performed recalling
that the Compton wavelength $\lambda_C=h/(mc)$ of a particle (in km)
is related to its mass (in eV) by
\be\label{lambdaC}
\lambda_C[{\rm km}]\times m[{\rm eV}]=1.24\times 10^{-9}\,.
\ee
Note also that a ten-Solar Mass black hole has a Schwarzschild radius
of $\sim 30~{\rm km}$. Compton wavelengths comparable to the size of
astrophysical black holes correspond to very light particles of mass
$\approx 10^{-10}$~eV (this observation will be useful in Chapter
\ref{ch:applications}).  Using a standard identity from tensor
analysis, the Klein-Gordon equation can be written
\be \frac{1}{\sqrt{-g}}\partial_{\mu}\left
(\sqrt{-g}g^{\mu\nu}\partial_{\nu}\Phi\right )-\mu^2\Phi=0\,,\label{kg2} \ee
where all quantities refer to the metric (\ref{Sch}). Since the metric
is spherically symmetric, the field evolution should be independent of
rotations: this suggests that the angular variables $\theta$ and
$\phi$ should factor out of the problem. The field can be decomposed
as
\be \Phi(t,r,\theta,\phi)=
\sum_{l=0}^\infty \sum_{m=-l}^l \frac{\Psi^{s=0}_{lm}(r)}{r}P_{lm}(\theta)\,
e^{-i\omega t}\, e^{im\phi}\,,
\label{separationscalar} \ee
where the superscript ``$s=0$'' is a reminder that these are spin-zero
perturbations, $Y_{lm}(\theta,\phi)=P_{lm}(\theta)\,e^{im\phi}$ are
the usual scalar spherical harmonics, and the associated Legendre
functions $P_{lm}(\theta)$ satisfy
\be \frac{1}{\sin\theta}\partial_{\theta}\left (\sin\theta
\partial_{\theta}P_{lm}\right
)-\frac{m^2}{\sin^2\theta}P_{lm}=-l(l+1)P_{lm}\,.\label{sphe} \ee
Note that this decomposition is the most general possible, because the
spherical harmonics are a complete orthonormal set of functions.
Inserting (\ref{separationscalar}) into equation (\ref{kg2}) and using
(\ref{sphe}), we get the following radial wave equation for
$\Psi^{s=0}_{lm}(r)$:
\be
fh \frac{d^2\Psi^{s=0}_{l}}{dr^2}+\f{1}{2}(fh)'\frac{d\Psi^{s=0}_{l}}{dr}+
\left\lbrack\omega^2-
\left(\mu^2 f+\frac{l(l+1)}{r^2} f-\frac{(fh)'}{2r}\right
)\right\rbrack\Psi^{s=0}_{l}=0\,,
\ee
where we can write $\Psi^{s=0}_{l}(r)$ rather than
$\Psi^{s=0}_{lm}(r)$ because the differential equation does not depend
on $m$ (the background is spherically symmetric).  In analogy with
(\ref{tortoise}), we can introduce the ``generalized tortoise
coordinate''
\be
\f{dr}{dr_*}\equiv (fh)^{1/2}
\ee
to eliminate the first-order derivative and obtain the following
radial equation:
\be
\frac{d^2\Psi^{s=0}_{l}}{dr_*^2}+
\left\lbrack\omega^2-
V_0(\mu)
\right\rbrack\Psi^{s=0}_{l}=0\,, \label{waveeqscalar} 
\ee
where the mass-dependent scalar potential reads
\be
V_0(\mu)\equiv f \mu^2+f \frac{l(l+1)}{r^2} - \frac{(fh)'}{2r}\,.
\ee
The tortoise coordinate will turn out to be useful because, in the
Schwarzschild metric, the location of the horizon ($r=2M$) corresponds
to $r_*\to -\infty$, so the problem of black-hole perturbations can be
formulated as a one-dimensional scattering problem over the real axis.

\begin{figure*}[t]
\begin{center}
\begin{tabular}{ll}
\epsfig{file=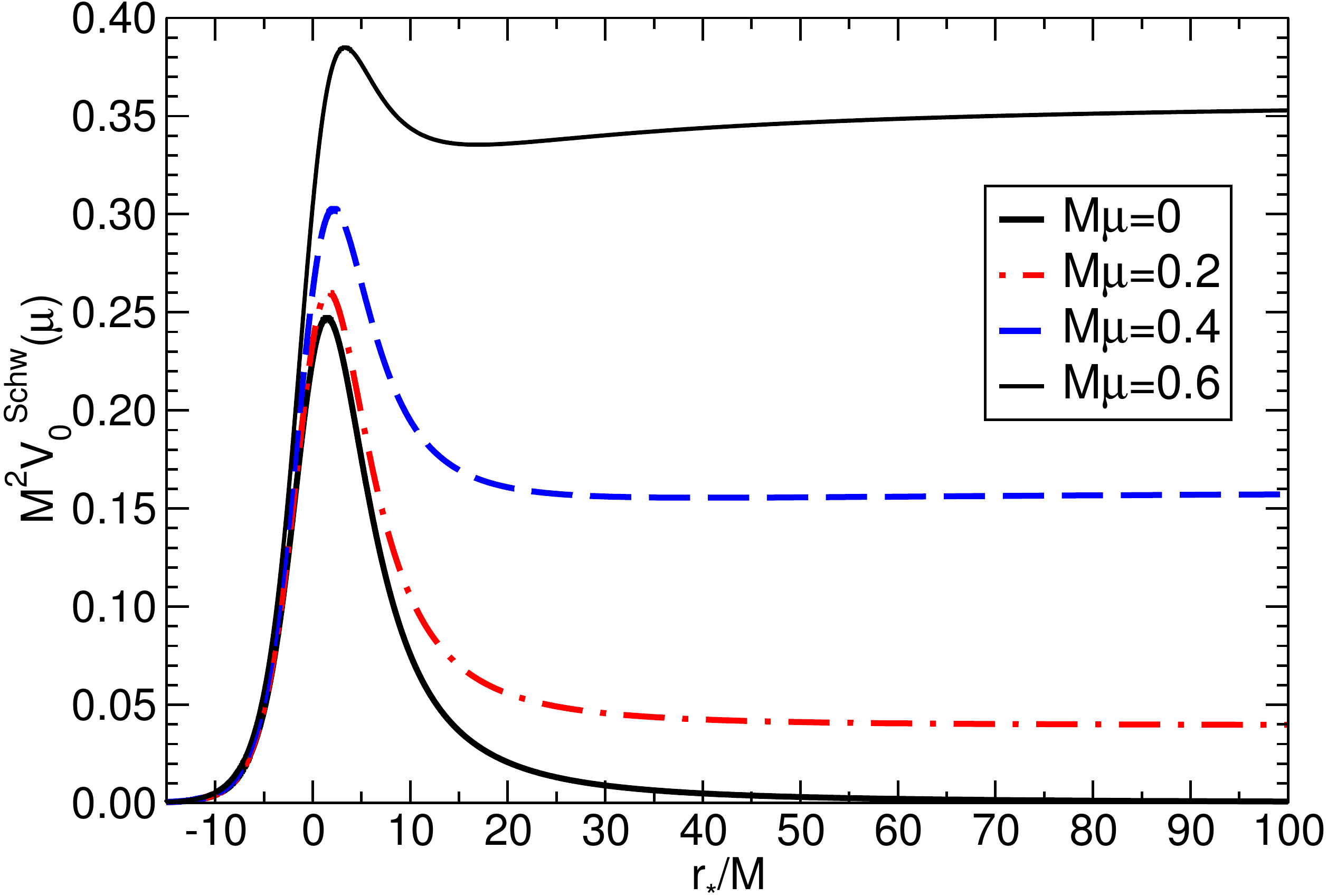,width=\textwidth,angle=0,clip=true}&
\end{tabular}
\caption{Schwarzschild potential (\ref{waveeqscalar-Sch}) for massive
  scalar perturbations with $l=2$ and different values of the scalar
  mass ($\mu=0,\,0.2,\,0.4,\,0.6$). The mass term acts as a confining
  potential: it gives rise to a local minimum and allows for the
  existence of quasi-bound states.
\label{fig:SchwPot}}
\end{center}
\end{figure*}

For most black-hole metrics of interest, $f=h$. In particular, for the
Schwarzschild metric the potential reads
\be
V_0^{\rm Schw}(\mu)=\left(1-\f{2M}{r}\right) 
\left(\mu^2+\frac{l(l+1)}{r^2} + \f{2M}{r^3}\right)\,,
\label{waveeqscalar-Sch} 
\ee
and the tortoise coordinate satisfies
\be
\f{dr_*}{dr}=\left(1-\f{2M}{r}\right)^{-1}=1+\f{2M}{r-2M}\,,
\ee
which is easily integrated to give
\be
r_*=r+2M \ln\left(\f{r}{2M}-1\right)\,,
\ee
where the integration constant has been fixed using a common choice in
the literature. The potential (\ref{waveeqscalar-Sch}) has some
crucial properties that are illustrated in Fig.~\ref{fig:SchwPot}: (i)
it tends to zero as $r\to 2M$ ($r_*/M\to -\infty$), where $f=1-2M/r\to
0$; (ii) it tends to $\mu^2$ as $r/M\to \infty$ ($r_*/M\to \infty$),
so whenever $\mu\neq 0$ there is a ``reflective potential barrier'' at
infinity; (iii) when the mass term is nonzero there is a local minimum
outside the peak of the potential, that allows for the existence of
quasi-bound states (that slowly leak out to infinity because of
dissipation).

\index{Master Equation!Schwarzschild|(}
\begin{BOX}{Master Equation for Massless Bosonic Fields, and the Peak of the Potential}
\label{box:MasterEq}
Separating the equations describing massless spin-one and spin-two
(i.e., electromagnetic and gravitational) perturbations in the
Schwarzschild background is not as easy as separating the angular
dependence for the scalar field (things get even harder for massive
fields: see references in Section \ref{sec:BHBombs} below). The
angular separation can be performed using vector and tensor spherical
harmonics (see e.g. \cite{Ruffini,Zerilli:1971wd}), and it leads to a
remarkable result: perturbations induced by any of these bosonic
fields satisfy a single ``master equation'',
\be
\frac{d^2\Psi^{s}_{l}}{dr_*^2}+
\left\lbrack \omega^2-V_{s} \right\rbrack
\Psi^{s}_{l}=0\,, \label{mastereq} 
\ee
where
\beq
V_s&=&f \left(\frac{l(l+1)}{r^2} + \f{2M (1-s^2)}{r^3}\right)\nn\\
&=&f \left(\f{\Lambda}{r^2} + \f{2\beta}{r^3}\right)
\,.
\eeq
In the last line we set $M=1$ to simplify the algebra. We also
introduced $\beta\equiv 1-s^2$ and $\Lambda=l(l+1)$, where
$s=0,\,1,\,2$ is the spin of the perturbing field. Then
Eq.~(\ref{waveeqscalar-Sch}) is just a special case for $s=0$ and
$\mu=0$.
Let us look for the extrema of the function $Q(r_*)=\omega^2-V_s(r)$.
It is easy to show that
\be
Q'(r_*)=
\left(1-\f{2}{r}\right)
\left\{
\f{\Lambda}{r^3}\left(2-\f{6}{r}\right)+
\f{2\beta}{r^4}\left(3-\f{8}{r}\right)
\right\}
\ee
Besides the asymptote as $r\to 2$, the only zero occurs when the curly
brace vanishes. Multiplying the curly brace by $r^5$ one gets a
quadratic equation, that is easily solved for the location of the
extremum:
\be
r_0=\f{3}{2\Lambda}
\left\{
\Lambda-\beta+\sqrt{\Lambda^2+\beta^2+\f{14}{9}\Lambda\beta}
\right\}\,.
\ee
It is readily seen that $r_0\to 3$ when $\Lambda\to \infty$ (i.e. in
the eikonal, or geometric optics, limit). Even for small $l$'s the
location of the peak is very close to the marginal photon orbit:
for $l=2$, $s=(0,\,1,\,2)$ one gets $r_0=(2.95171,\,3,\,3.28078)$; 
for $l=3$, $s=(0,\,1,\,2)$ one gets $r_0=(2.97415,\,3,\,3.0001)$.
\end{BOX}
\index{Master Equation!Schwarzschild|(}

\section{Solution of the Scattering Problem}

Unlike most idealized macroscopic physical systems, perturbed
black-hole spacetimes are intrinsically dissipative due to the
presence of the event horizon, which acts as a one-way membrane. In
fact, the system is ``leaky'' both at the horizon and at infinity,
where energy is dissipated in the form of gravitational
radiation. This precludes a standard normal-mode analysis: dissipation
means that the system is not time-symmetric, and that the associated
boundary-value problem is non-Hermitian. In general, after a transient
that depends on the source of the perturbation, the response of a
black hole is dominated by characteristic complex frequencies (the
``quasinormal modes'', henceforth QNMs). The imaginary part is nonzero
because of dissipation, and its inverse is the decay timescale of the
perturbation. The corresponding eigenfunctions are usually not
normalizable, and, in general, they do not form a complete set
(cf. \cite{RevModPhys.70.1545,Nollert:1998ys} for more extensive
discussions). Almost any real-world physical system is dissipative, so
one might reasonably expect QNMs to be ubiquitous in physics. QNMs are
indeed useful in the treatment of many dissipative systems, e.g. in
the context of atmospheric science and leaky resonant
cavities. Extensive reviews on black-hole QNMs and their role in
various fields of physics (including gravitational-wave astronomy, the
gauge-gravity duality and high-energy physics) can be found in
\cite{Nollert:1999ji,Kokkotas:1999bd,Berti:2009kk,Konoplya:2011qq}.

Black-hole QNMs were initially studied to assess the stability of the
Schwarzschild metric \cite{Vishveshwara:1970cc,pressringdown}. Goebel
\cite{goebel} was the first to realize that they can be thought of as
waves traveling around the black hole: more precisely, they can be
interpreted as waves {\em trapped at the unstable circular null
  geodesic} (the ``light ring'') and slowly leaking out. This
qualitative picture was refined by several authors over the years
\cite{Ferrari:1984zz,mashhoon,stewart,Decanini:2002ha,
  Cardoso:2008bp,Decanini:2009mu,Dolan:2009nk}. It will be shown below
that the instability timescale of the geodesics is the decay timescale
of the QNM, and the oscillation frequency $\omega\sim c/r_{\rm mb}$,
with $c$ the speed of light and $r_{\rm mb}$ the light-ring radius
(see \cite{Cardoso:2008bp} for generalizations of these arguments to
rotating and higher-dimensional black holes).

One of the most fascinating aspects of black-hole physics is that the
master equation (\ref{mastereq}) can be solved using methods familiar
from nonrelativistic quantum mechanics, in particular from scattering
theory.  We will first review a method developed by Leaver
\cite{Leaver:1985ax} to compute QNM frequencies ``exactly'' (within
the limits of a computer's numerical accuracy). Leaver's method
follows quite closely techniques that were developed to deal with the
hydrogen ion in quantum mechanics as early as 1934 \cite{Jaffe:1934}.
Then we will confirm the intuitive picture of black-hole QNMs as
light-ring perturbation using one of the simplest ``textbook''
approximation technique to solve scattering problems: the
Wentzel-Kramers-Brillouin (WKB) approximation. In the black-hole
perturbation theory context, the WKB approximation was first
investigated by Mashhoon \cite{Mashhoon:1985}, Schutz and Will
\cite{Schutz:1985km}.

\subsection{Leaver's Solution}\label{sec:leaver}

Leaver's solution of the wave equation is based on a classic 1934
paper by Jaff\'e on the electronic spectra of the hydrogen molecular
ion \cite{Jaffe:1934} -- quantum mechanics comes to the rescue again!
The wave equation is found to be a special case of the so-called
``generalized spheroidal wave equation'' \cite{leJMP}, for which the
solution can be written as a series expansion. By replacing the series
expansion into the differential equations and imposing QNM boundary
conditions, one finds recursion relations for the expansion
coefficients.  The series solution converges only when a certain
continued fraction relation involving the mode frequency and the black
hole parameters holds. The evaluation of continued fractions amounts
to elementary algebraic operations, and the convergence of the method
is excellent, even at high damping. In the Schwarzschild case, in
particular, the method can be tweaked to allow for the determination
of modes of order up to $\sim 100,000$ \cite{Nollert:1993zz}.

As stated (without proof) in Box~\ref{box:MasterEq}, the angular
dependence of the metric perturbations of a Schwarzschild black hole
can be separated using tensorial spherical harmonics
\cite{Zerilli:1971wd}. Depending on their behavior under parity
transformations, the perturbation variables are classified as {\it
  polar} (even) or {\it axial} (odd). The resulting differential
equations can be manipulated to yield two wave equations, one for the
polar perturbations (that we shall denote by a ``plus'' superscript)
and one for the axial perturbations (``minus'' superscript):
\be\label{wave-nonr}
\left({d^2\over dr_*^2}+\omega^2\right)Z_l^{\pm}=V_l^{\pm} Z_l^{\pm}\,.
\ee
We are interested in solutions of equation (\ref{wave-nonr}) that are
purely outgoing at spatial infinity ($r\to \infty$) and purely ingoing
at the black hole horizon ($r\to 1$). These boundary conditions are
satisfied by an infinite, discrete set of complex frequencies
$\omega=\omega_R+\ii \omega_I$ (the QNM frequencies). For the
Schwarzschild solution the two potentials $V^{\pm}$ are quite
different, yet the QNMs for polar and axial perturbations are the
same. The proof of this suprising fact can be found in \cite{MTB}. The
underlying reason is that polar and axial perturbations are related by
a differential transformation discovered by Chandrasekhar \cite{MTB},
and both potentials can be seen to emerge from a single
``superpotential''. In fancier language, the axial and polar
potentials are related by a supersymmetry transformation (where
``supersymmetry'' is to be understood in the sense of nonrelativistic
quantum mechanics): cf. \cite{Leung:1999fr}. Since polar and axial
perturbations are isospectral and $V^-$ has an analytic expression
which is simpler to handle, we can focus on the axial equation for
$Z_l^{-}=\Psi^{s=2}_{l}$, i.e. the master equation (\ref{mastereq})
with $s=2$.

We will now find an algebraic relation that can be solved
(numerically) to determine the eigenfrequencies of scalar,
electromagnetic and gravitational perturbations of a Schwarzschild
black hole. For a study of the differential equation it is convenient
to use units where $2M=1$, so that the horizon is located at $r=1$ (we
will only use these units in this section). The tortoise coordinate
and the usual Schwarzschild coordinate radius $r$ are related by
\be
\f{dr}{dr_*}=\f{\Delta}{r^2}\,,
\ee
with $\Delta=r(r-1)$.  When written in terms of the ``standard''
radial coordinate $r$ and in units $2M=1$, the master equation
(\ref{mastereq}) reads:
\be\label{sax}
r(r-1)\frac{d^2\Psi^{s}_{l}}{dr^2}+\frac{d\Psi^{s}_{l}}{dr}-
\left[l(l+1)-\frac{s^2-1}{r}-\frac{\omega^2 r^3}{r-1}\right]\Psi^{s}_{l}=0\,,
\ee
where $s$ is the spin of the perturbing field ($s=0,~1,~2$ for scalar,
electromagnetic and gravitational perturbations, respectively) and $l$
is the angular index of the perturbation. 
Perturbations of a Schwarzschild background are independent of the
azimuthal quantum number $m$, because of spherical symmetry (this is
not true for Kerr black holes).
Equation (\ref{sax}) can be solved using a series expansion of
the form:
\be\label{sser}
\Psi^{s}_{l}=(r-1)^{-\ii \omega}r^{2\ii \omega}e^{\ii \omega (r-1)}
\sum_{j=0}^{\infty} a_j \left(\frac{r-1}{r}\right)^j\,,
\ee
where the prefactor is chosen to incorporate the QNM boundary
conditions at the horizon and at infinity:
\beq
&&\Psi^{s}_{l}\sim e^{\ii \omega r_*}
\sim e^{\ii \omega (r+\ln r)} 
\sim r^{\ii \omega} e^{\ii \omega r} 
\quad {\rm as} \quad r\to \infty\,,\\
&&\Psi^{s}_{l}\sim e^{-\ii \omega r_*}
\sim e^{-\ii \omega \ln (r-1)}
\sim (r-1)^{-\ii \omega}
\quad {\rm as} \quad r\to 1\,.
\eeq
Substituting the series expansion (\ref{sser}) in (\ref{sax}) we get a
three term recursion relation for the expansion coefficients $a_j$:
\beq
&&\alpha_0 a_1+\beta_0 a_0=0,\\
&&\alpha_n a_{n+1}+\beta_n a_n+\gamma_n a_{n-1}=0,
\qquad n=1,2,\dots\nn
\eeq
where $\alpha_j$, $\beta_j$ and $\gamma_j$ are simple functions of the
frequency $\omega$, $l$ and $s$ \cite{Leaver:1985ax}:
\beq
&&\alpha_n=n^2+(2-2\ii\omega)n+1-2\ii\omega\,,\\
&&\beta_n=-\left[2n^2+(2-8\ii\omega)n-8\omega^2-4\ii\omega+l(l+1)+1-s^2\right]\,,\\
&&\gamma_n=n^2-4\ii\omega n-4\omega^2-s^2\,.
\eeq
A mathematical theorem due to Pincherle guarantees that the series is
convergent (and the QNM boundary conditions are satisfied) when the
following continued fraction condition on the recursion coefficients
holds:
\be\label{CF}
0=\beta_0-
{\alpha_0\gamma_1\over \beta_1-}
{\alpha_1\gamma_2\over \beta_2-}\dots
\ee
The $n$--th QNM frequency is (numerically) the most stable root of the
$n$--th inversion of the continued-fraction relation (\ref{CF}), i.e.,
it is a solution of
\be\label{CFI}
\beta_n-
{\alpha_{n-1}\gamma_{n}\over \beta_{n-1}-}
{\alpha_{n-2}\gamma_{n-1}\over \beta_{n-2}-}\dots
{\alpha_{0}\gamma_{1}\over \beta_{0}}=
{\alpha_n\gamma_{n+1}\over \beta_{n+1}-}
{\alpha_{n+1}\gamma_{n+2}\over \beta_{n+2}-}\dots
\qquad (n=1,2,\dots).\nn
\ee
The infinite continued fraction appearing in equation (\ref{CFI}) can
be summed ``bottom to top'' starting from some large truncation index
$N$. Nollert \cite{Nollert:1993zz} has shown that the convergence of the procedure
improves if the sum is started using a wise choice for the value of
the ``rest'' of the continued fraction, $R_N$, defined by the relation
\be
R_N={\gamma_{N+1}\over\beta_{N+1}-\alpha_{N+1}R_{N+1}}\,.
\ee
Assuming that the rest can be expanded in a series of the form
\be\label{RN}
R_N=\sum_{k=0}^{\infty}C_k N^{-k/2}\,,
\ee
it turns out that the first few coefficients in the series are
$C_0=-1$, $C_1=\pm\sqrt{-2\ii \omega}$, $C_2=(3/4+2\ii \omega)$ and
$C_3=\left[l(l+1)/2+2\omega^2+3\ii\omega/2+3/32\right]/C_1$ (the
latter coefficient contains a typo in \cite{Nollert:1993zz}, but this
is irrelevant for numerical calculations).

\begin{figure*}[thb]
\begin{center}
\epsfig{file=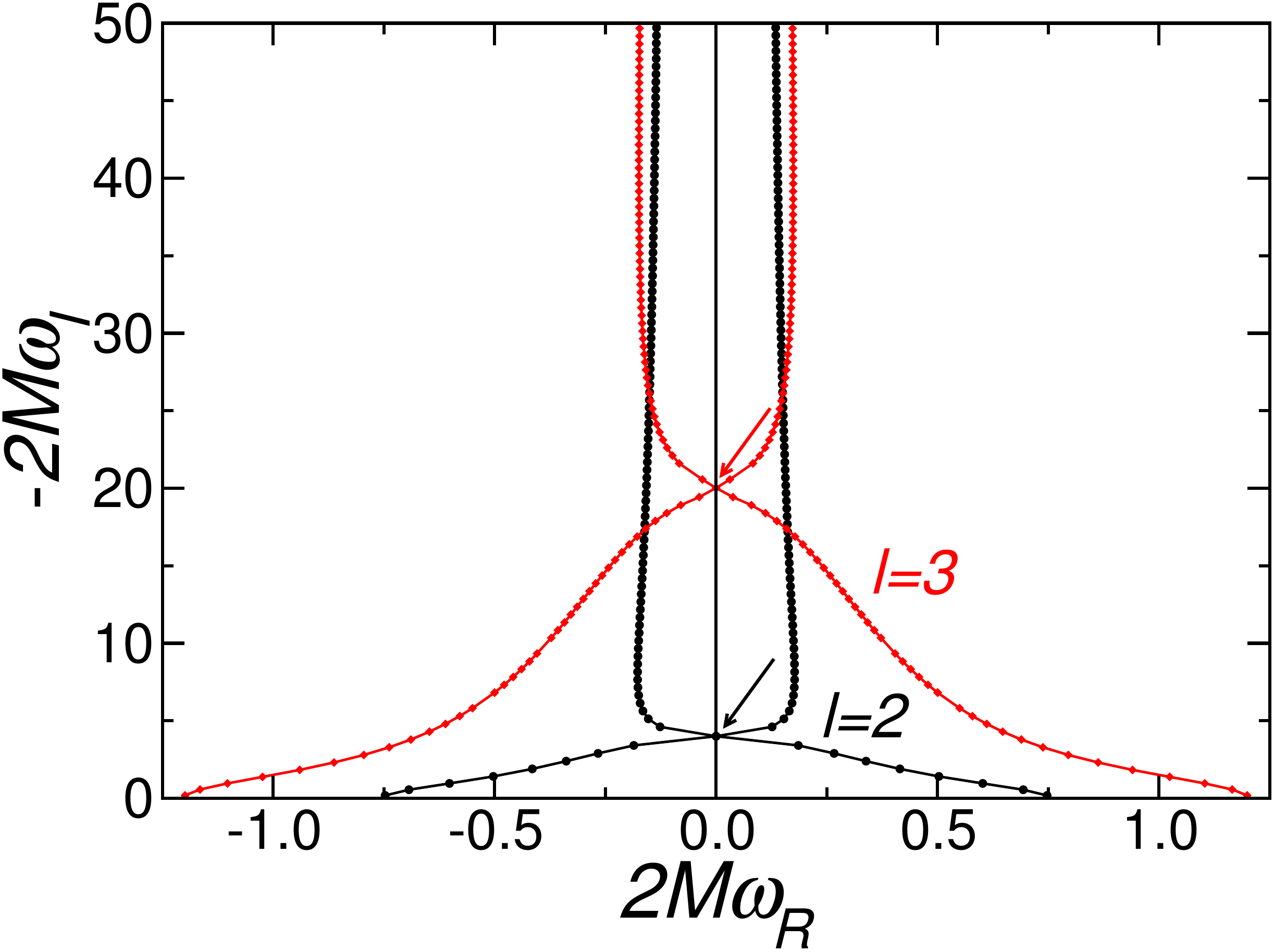,width=\textwidth,angle=0}
\caption{Quasinormal frequencies for gravitational perturbations with
  $l=2$ (blue circles) and $l=3$ (red diamonds). Compare eg. Figure 1
  in \cite{Nollert:1993zz}. In both cases we mark by an arrow the
  algebraically special mode, that is given analytically by Equation
  (\ref{AlgSp}). Notice that as the imaginary part of the frequency
  tends to infinity the real part tends to a finite, {\it
    $l$-independent} limit. (From Ref.~\cite{Berti:2009kk}.)
\label{fig:fig1}}
\end{center}
\end{figure*}

The dominant QNM frequencies with $l=2$ and $l=3$ resulting from this
procedure are shown in Table \ref{tab1} and Figure
\ref{fig:fig1}. Reintroducing physical units, the fundamental
oscillation frequency $f=\omega_R/(2\pi)$ and the damping time
$\tau=1/|\omega_I|$ of an astrophysical black hole scale with mass
according to the relation
\be
f=1.207 \left(\frac{10~M_\odot}{M}\right) {\rm kHz}\,,\qquad
\tau=0.5537 \left(\frac{M}{10~M_\odot}\right) {\rm ms}\,.
\ee
An ``algebraically special'' mode, whose frequency is (almost) purely
imaginary, separates the lower QNM branch from the upper branch
\cite{chandraspecial}. This algebraically special mode is located at
\be\label{AlgSp}
\tilde \Omega_l=\pm \ii{(l-1)l(l+1)(l+2)\over 6}\,,
\ee
and it can be taken as roughly marking the onset of the asymptotic
high-damping regime. The algebraically special mode quickly moves
downwards in the complex plane (i.e., upwards in the figure) as $l$
increases: from Table \ref{tab1} we see that it corresponds to an
overtone index $n=9$ when $l=2$, and to an overtone index $n=41$ when
$l=3$. This means that for high values of $l$ the asymptotic
high-damping regime sets in later, becoming harder to probe using
numerical methods.
The algebraically special modes are a fascinating technical subject:
they can be shown to be related to the instability properties of naked
singularities \cite{Cardoso:2006bv}, and (strictly speaking) they are
not even QNMs, because they do not satisify QNM boundary conditions:
see e.g.~\cite{Berti:2004md} and references therein.

\begin{table}[htb]
\centering
\caption{Representative Schwarzschild quasinormal frequencies for
  $l=2$ and $l=3$ (from \cite{Leaver:1985ax}).}
\vskip 12pt
\begin{tabular}{@{}lll@{}}
\hline
\hline
  &$l=2$                 & $l=3$     \\
\hline
n &$2M\omega_n$            & $2M\omega_n$\\
\hline
1 &(0.747343,-0.177925)  &(1.198887,-0.185406)\\
2 &(0.693422,-0.547830)  &(1.165288,-0.562596)\\
3 &(0.602107,-0.956554)	 &(1.103370,-0.958186)\\
4 &(0.503010,-1.410296)	 &(1.023924,-1.380674)\\
5 &(0.415029,-1.893690)	 &(0.940348,-1.831299)\\
6 &(0.338599,-2.391216)	 &(0.862773,-2.304303)\\
7 &(0.266505,-2.895822)	 &(0.795319,-2.791824)\\
8 &(0.185617,-3.407676)	 &(0.737985,-3.287689)\\
9 &(0.000000,-3.998000)	 &(0.689237,-3.788066)\\
10&(0.126527,-4.605289)	 &(0.647366,-4.290798)\\
11&(0.153107,-5.121653)	 &(0.610922,-4.794709)\\
12&(0.165196,-5.630885)	 &(0.578768,-5.299159)\\
20&(0.175608,-9.660879)	 &(0.404157,-9.333121)\\
30&(0.165814,-14.677118) &(0.257431,-14.363580)\\
40&(0.156368,-19.684873) &(0.075298,-19.415545)\\
41&(0.154912,-20.188298) &(-0.000259,-20.015653)\\
42&(0.156392,-20.685630) &(0.017662,-20.566075)\\
50&(0.151216,-24.693716) &(0.134153,-24.119329)\\
60&(0.148484,-29.696417) &(0.163614,-29.135345)\\
\hline
\hline
\end{tabular}
\label{tab1}
\end{table}

Nollert was the first to compute highly damped quasinormal frequencies
corresponding to {\it gravitational} perturbations
\cite{Nollert:1993zz}. His main result was that the real parts of the
quasinormal frequencies are well fitted, for large $n$, by a relation
of the form
\be\label{fit}
\omega_R=\omega_{\infty}+{\lambda_{s,l}\over \sqrt{n}}\,.
\ee
These numerical results are perfectly consistent with analytical
calculations \cite{Berti:2003zu}. Motl \cite{Motl:2002hd} analyzed the
continued fraction condition (\ref{CF}) to find that highly damped
quasinormal frequencies satisfy the relation
\be\label{Mresult}
\omega\sim{T_H \ln 3}+(2n+1)\pi \ii T_H+{\cal O}(n^{-1/2})\,.
\ee
(in units $2M=1$, the Hawking temperature of a Schwarzschild black
hole $T_H=1/4\pi$). This conclusion was later confirmed by
complex-integration techniques \cite{Motl:2003cd} and phase-integral
methods \cite{Andersson:2003fh}, and it may have a connection with
Bekenstein's ideas on black-hole area quantization \cite{Hod:1998vk}.

\subsection{The WKB Approximation}

While Leaver's method provides the most accurate numerical solution of
the scattering problem, the WKB approximation is useful to develop
physical intuition on the meaning of QNMs. The derivation below
follows quite closely the paper by Schutz and Will
\cite{Schutz:1985km}, which in turn is based on the excellent
treatment in the book by Bender and Orszag
\cite{benderorszag}. Consider the equation\footnote{In quantum mechanics, the Schr\"odinger equation for a
particle of mass $m$ and energy $E$ moving in a one-dimensional
potential $V(x)$
\be
\left[-\f{\hbar^2}{2m}\f{d^2}{dx^2}+V(x)\right]\Psi=E\Psi
\ee
can be rewritten in the previous form with
\be
-Q(x)=\f{2m}{\hbar^2}\left[V(x)-E\right]\,.
\ee}
\be\label{WEWKB}
\f{d^2\Psi}{dx^2}+Q(x)\Psi=0\,,
\ee
where $\Psi$ is the radial part of the perturbation variable (the time
dependence has been separated by Fourier decomposition, and the
angular dependence is separated using the scalar, vector or tensor
spherical harmonics appropriate to the problem at hand). The
coordinate $x$ is a tortoise coordinate such that $x\to -\infty$ at
the horizon and $x\to \infty$ at spatial infinity. The function
$-Q(x)$ is constant in both limits ($x\to \pm\infty$) but not
necessarily the same at both ends, and it rises to a maximum in the
vicinity of $x=0$ (more specifically, as we have seen above, at
$r\simeq 3M$). 
Since $Q(x)$ tends to a constant at large $|x|$, we have
\be
\Psi\sim e^{\pm \ii \alpha x}\,, \quad {\rm as} \quad x\to \pm \infty\,
\ee
with ${\rm Re}(\alpha)>0$. If $Q(x)\to 0$, $\omega=\alpha$. With our
convention on the Fourier decomposition ($\Psi\sim e^{-\ii \omega
  t}$), outgoing waves at $x\to\infty$ correspond to the positive sign
in the equation above, and waves going into the horizon as $x\to
-\infty$ correspond to the negative sign in the equation above
(cf.~Fig.~\ref{fig:SchwPot}).

The domain of definition of $Q(x)$ can be split in three regions: a
region I to the left of the turning point where $Q(x_{\rm I})=0$, a
``matching region'' II with $x_{\rm I}<x<x_{\rm II}$, and a region III
to the right of the turning point where $Q(x_{\rm II})=0$.
As shown in Box \ref{box:WKB}, in regions I and III we can write the
solution in the ``physical optics'' WKB approximation:
\beq
&&\Psi_{\rm I}\sim Q^{-1/4}\exp\left\{
\pm{\ii}\int_{x_2}^x [Q(t)]^{1/2}dt
\right\}\,,\\
&&\Psi_{\rm III}\sim Q^{-1/4}\exp\left\{
\pm{\ii}\int_{x}^{x_1} [Q(t)]^{1/2}dt
\right\}\,,
\eeq

\index{WKB!Bender|(}
\begin{BOX}{WKB Approximation: ``Physical Optics''}
\label{box:WKB}
Introduce a perturbative parameter $\epsilon$ (proportional to $\hbar$
in quantum mechanics) and write the ODE as
\be
\epsilon^2 \Psi''=Q(x)\Psi\,.
\ee
Now write the solution in the form
\be
\Psi(x)\sim 
\exp\left[ \f{1}{\epsilon}\sum_{n=0}^\infty \epsilon^n S_n(x) \right]
\,,
\ee
and compute the derivatives:
\beq
&&\Psi'\sim 
\left(\f{1}{\epsilon} \sum_{n=0}^\infty \epsilon^n S'_n\right)
\exp\left[ \f{1}{\epsilon}\sum_{n=0}^\infty \epsilon^n S_n\right]
\,,\\
&&\Psi''\sim 
\left[
\f{1}{\epsilon^2} \left( 
\sum_{n=0}^\infty \epsilon^n S'_n
\right)^2
+
\f{1}{\epsilon} \sum_{n=0}^\infty \epsilon^n S''_n
\right]
\exp\left[ \f{1}{\epsilon}\sum_{n=0}^\infty \epsilon^n S_n \right]
\,,\\
\eeq
Substituting into the ODE and dividing by the common exponential
factors we get
\be
(S_0')^2+2\epsilon S_0' S_1' +\epsilon S_0''+\dots = Q(x)\,.
\ee
The first two terms in the expansion yield the so-called ``eikonal
equation'' and ``transport equation'':
\beq
&&(S_0')^2=Q(x)\,,\\
&&2S_0'S_1'+S_0''=0\,.
\eeq
The solution to the eikonal equation is
\be
S_0(x)=\int^x \sqrt{Q(t)}dt\,,
\ee
while the solution to the transport equation is
\be
S_1(x)=-\f{1}{4}\ln Q(x)+{\rm constant}\,.
\ee
The leading-order solution is called the ``geometrical optics''
approximation. The next-to-leading order solution (including $S_1(x)$)
is called the ``physical optics'' approximation. For higher-order
solutions, cf.~\cite{benderorszag,Iyer:1986np,Konoplya:2003ii}.
\end{BOX}
\index{WKB!Bender|(}

\begin{figure*}[t]
\begin{center}
\begin{tabular}{ll}
\epsfig{file=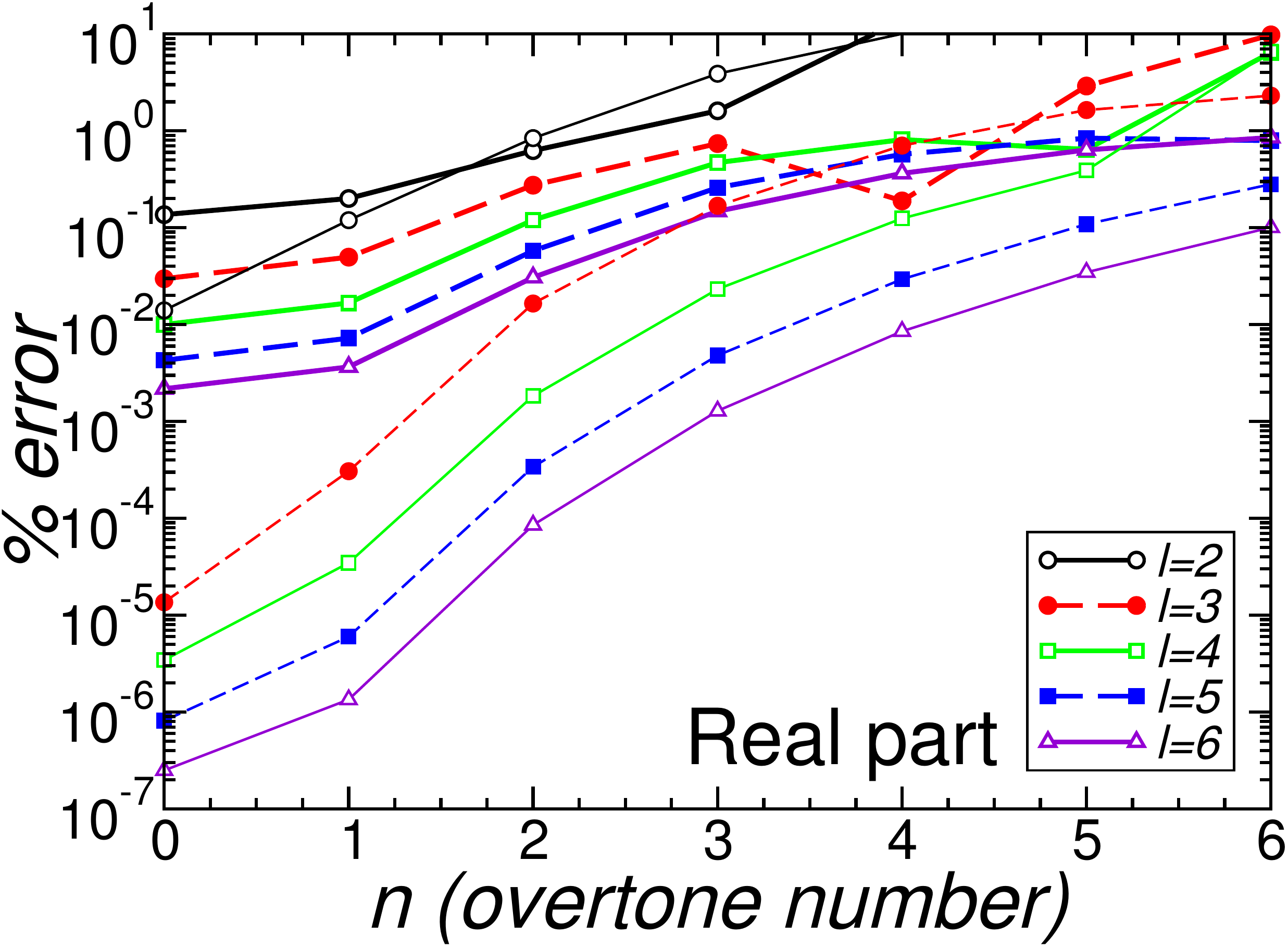,width=0.5\textwidth,angle=0,clip=true}&
\epsfig{file=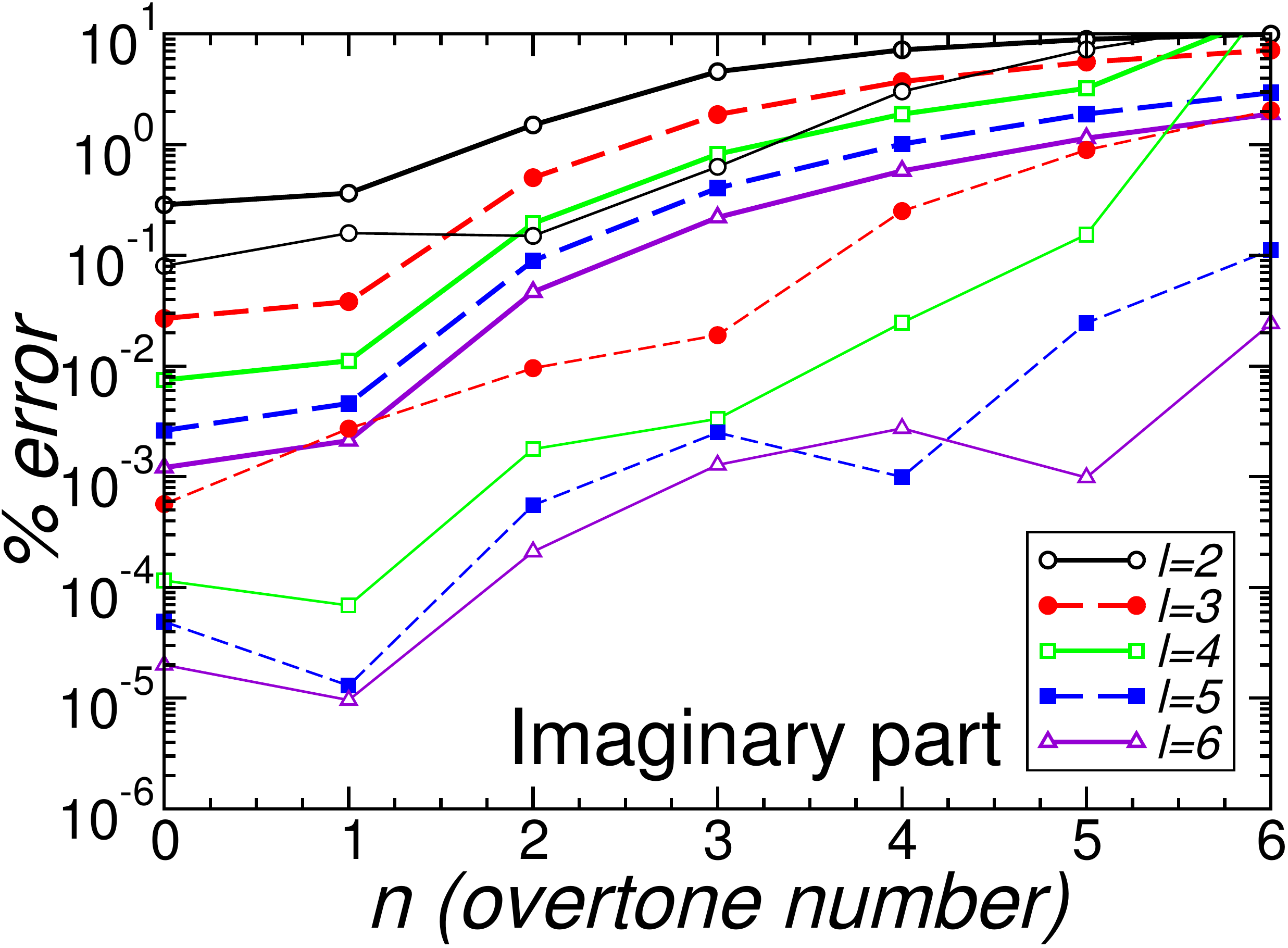,width=0.5\textwidth,angle=0,clip=true}
\end{tabular}
\caption{Accuracy of the WKB approximation for the real part (left
  panel) and imaginary part (right panel) of the QNM
  frequencies. (From Ref.~\cite{Berti:2009kk}.)
\label{fig:WKB}}
\end{center}
\end{figure*}

The idea is now to find the equivalent of the Bohr-Sommerfeld
quantization rule from quantum mechanics. We want to relate two WKB
solutions across the ``matching region'' whose limits are the
classical turning points, where $\omega^2=V(r)$. The technique works
best when the classical turning points are close, i.e. when $\omega^2
\sim V_{\rm max}$, where $V_{\rm max}$ is the peak of the potential.
In region II, we expand
\be
Q(x)=Q_0+\f{1}{2}Q_0''(x-x_0)^2+{\cal O}(x-x_0)^3\,.
\ee
If we set $k\equiv Q_0''/2$ and we make the change of variables
%
$t=(4k)^{1/4}e^{\ii \pi/4}(x-x_0)$,
%
we find that the equation reduces to
\be
\f{d^2\Psi}{dt^2}+\left[
-\f{\ii Q_0}{(2Q_0'')^{1/2}}
-\f{1}{4}t^2\right]\Psi=0\,.
\ee
We now define a parameter $\nu$ such that
\be
\nu+\f{1}{2}=-\f{\ii Q_0}{(2Q_0'')^{1/2}}\,.
\ee
Then the differential equation takes the form
\be
\f{d^2\Psi}{dt^2}+\left[
\nu+\f{1}{2}
-\f{1}{4}t^2\right]\Psi=0\,,
\ee
whose solutions are parabolic cylinder functions, commonly denoted by
$D_{\nu}(z)$ \cite{Abramowitz:1970as,benderorszag}. These special
functions are close relatives of Hermite polynomials (this makes
sense, because the quadratic approximation means that we locally
approximate the potential as a quantum harmonic oscillator problem,
and Hermite polynomials are well known to be the radial solutions of
the quantum harmonic oscillator). The solution in region II is
therefore a superposition of parabolic cylinder functions:
\be
\Psi = AD_{\nu}(z)
+ BD_{-\nu-1}(iz) \,,\quad z\equiv (2Q_0'')^{\frac{1}{4}}e^{i\frac{\pi}{4}}(r_*-\bar{r}_*)\,.
\ee
Using the asymptotic behavior of parabolic cylinder functions
\cite{Abramowitz:1970as,benderorszag} and imposing the outgoing-wave
condition at spatial infinity we get, near the horizon,
\be \Psi \sim  A e^{-i\pi
\nu}z^{\nu}e^{-z^2/4} -
i
\sqrt{2\pi}A
\left[\Gamma(-\nu)\right]^{-1}
e^{5i\pi/4}z^{-\nu-1}e^{z^2/4}
\,.\ee
QNM boundary conditions imply that the outgoing term, proportional to
$e^{z^2/4}$, should be absent, so $1/\Gamma(-\nu)=0$, or
$\nu=n(=0\,,1\,,2\,,...)$. Therefore the leading-order WKB
approximation yields a ``Bohr-Sommerfeld quantization rule'' defining
the QNM frequencies:
\be \label{BohrSommerfeld}
Q_0/\sqrt{2Q_0''}=i(n+1/2)\,,\quad n=0\,,1\,,2\,,...
\ee
Higher-order corrections to this parabolic approximation have been
computed \cite{pressringdown}. Iyer and Will
\cite{Iyer:1986np,Iyer:1986nq} carried out a third-order WKB
expansion, and Konoplya \cite{Konoplya:2003ii} pushed the expansion up
to sixth order. There is no rigorous proof of convergence, but the
results do improve for higher WKB orders. Fig.~\ref{fig:WKB} compares
numerical results for the QNMs of Schwarzschild black holes from
Leaver's continued fraction method against third-order (thick lines)
and sixth-order (thin lines) WKB predictions.  The WKB approximation
works best for low overtones, i.e. modes with a small imaginary part,
and in the eikonal limit of large $l$ (which corresponds to large
quality factors, or large $\omega_R/\omega_I$). The method assumes
that the potential has a single extremum, which is the case for most
(but not all) black-hole potentials: see e.g.~\cite{Ishibashi:2003ap}
for counterexamples.

%

\section{Geodesic Stability and Black-Hole Quasinormal Modes}

Let us reconsider the Bohr-Sommerfeld quantization condition
(\ref{BohrSommerfeld}). In a spherically symmetric, asymptotically
flat spacetime of the form (\ref{Sch}), the Klein-Gordon equation can
be written as in Eq.~(\ref{WEWKB}) with the tortoise coordinate
(\ref{tortoise}). In the eikonal limit ($l\rightarrow \infty$) we get
\be\label{Qeikonal}
Q\simeq \omega^2-f\frac{l^2}{r^2}\,.
\ee
It is easy to check that scalar, electromagnetic and gravitational
perturbations of static black holes have the same behavior in the
eikonal limit (in fact, the same conclusion applies to
higher-dimensional spacetimes
\cite{Kodama:2003jz,Ishibashi:2003ap,Kodama:2003kk}). In other words,
there is a well-defined geometric-optics (eikonal) limit where the
potential for a wide class of massless perturbations is
``universal''. For $Q$ in Eq.~(\ref{Qeikonal}) above we find that the
extremum of $Q$ satisfies $2f(r_0)=r_0f'(r_0)$, i.e. $r_0$ coincides
with the location of the null circular geodesic $r_0=r_c$, as given by
Eq.~(\ref{cirgeo}).  Furthermore, the WKB formula
(\ref{BohrSommerfeld}) implies that, in the large-$l$ limit,
\be
\omega_{\rm QNM}=l\sqrt{\frac{f_c}{r_c^2}}
-i\frac{(n+1/2)}{\sqrt{2}}
\sqrt{-\frac{r_c^2}{f_c}\,\left (\frac{d^2}{dr_*^2}\frac{f}{r^2}\right )_{r=r_c}}\,.
\ee
Comparing this result with Eqs.~(\ref{Omegacircular}) and
(\ref{lyaponov}), it follows that
\be\label{QNMLyapunov}
\omega_{\rm QNM}=\Omega_c\,l-i(n+1/2)\,|\lambda_0|\,.
\ee
This is one of the punchlines of these notes: {\bf in the eikonal
  approximation, the real and imaginary parts of the QNMs of any
  spherically symmetric, asymptotically flat spacetime are given by
  (multiples of) the frequency and instability timescale of the
  unstable circular null geodesics.}

Dolan and Ottewill~\cite{Dolan:2009nk} have introduced a WKB-inspired
asymptotic expansion of QNM frequencies and eigenfunctions in powers
of the angular momentum parameter $l+1/2$. Their asymptotic expansion
technique is easily iterated to high orders, and it is very accurate
(at least for spherically symmetric spacetimes).  The asymptotic
expansion also provides physical insight into the nature of QNMs,
nicely connecting the geometrical understanding of QNMs as
perturbations of unstable null geodesics with the singularity
structure of the Green's function.

\section{Superradiant Amplification}\label{subs:superradiance}

The discussion so far was limited to uncharged and/or nonrotating
black holes. The inclusion of charge and rotation leads to the
interesting possibility of superradiant amplification. I will first
explain the conceptual foundations of superradiance, and then look at
the astrophysically most interesting case of rotational superradiance
in the Kerr spacetime.  For a stationary, asymptotically flat black
hole, the equations describing spin-$s$ fields can be written in the
form
\be 
\frac{d^2\Psi}{dr_*^2}+V(\omega,\,r)\Psi=0\,,
\label{wave}
\ee
where as usual $\omega$ is the frequency in a Fourier transform with
respect to the asymptotic time coordinate: $\Psi(t)=e^{-i\omega
  t}\Psi(\omega)$, and the radius $r_*$ is a convenient tortoise
coordinate. As it turns out, the tortoise coordinate and the radial
potential at the horizon ($r\to r_+$) and at infinity ($r\to\infty$)
behave as follows:
\be
\left\{
\begin{array}{ll}
r_* \sim r \,,&   V \sim \omega ^2\,\,\, \,\,\,\,\quad
\quad {\rm as}\ r\rightarrow \infty \,, \\
e^{r_*} \sim (r-r_+)^{\alpha} \,,&   V \sim (\omega-\varpi)^2\,\,\,
{\rm as}\ r\rightarrow r_+ \,,
\end{array}
\right. \label{bound}
\ee
where $\alpha$ is a positive constant. The function $\varpi$ can be
related to rotation (in the Kerr geometry $\varpi=m\Omega$, with $m$ an
azimuthal number and $\Omega=a/(2Mr_+)$ the angular velocity of the
horizon) or it can be a ``chemical potential'' (in the
Reissner-Nordstr\"om geometry $\varpi=qQ$, where $q$ is the charge of
the perturbing field and $Q$ the charge of the black hole).

Waves scattering in this geometry have the following asymptotic
behavior:\footnote{For simplicity, here we consider a massless field,
  but the discussion is generalized to massive fields in a
  straightforward way.}
\be
\Psi _1 \sim\left\{
\begin{array}{ll}
\mathcal{T}_{\rm in}\,(r-r_+)^{-\ii \alpha (\omega-\varpi)}+\mathcal{T}_{\rm out}\,(r-r_+)^{\ii \alpha (\omega-\varpi)} & {\rm as}\
r\rightarrow r_+ \,,\label{bound2} \\
\mathcal{R}\,{\rm e}^{\ii {\omega r}}+ {\rm e}^{-\ii {\omega r}}&
{\rm as}\ r\rightarrow \infty\,.
\end{array}
\right.
\ee
These boundary conditions correspond to an incident wave of unit
amplitude coming from infinity, giving rise to a reflected wave of
amplitude $\mathcal{R}$ going back to infinity. At the horizon there
is a transmitted wave of amplitude $\mathcal{T}_{\rm in}$ going into
the horizon, and a wave of amplitude $\mathcal{T}_{\rm out}$ going out
of the horizon.

Assuming a real potential (this is true for massive scalar fields and
in most other cases), the complex conjugate of the solution $\Psi _1$
satisfying the boundary conditions (\ref{bound2}) will satisfy the
complex-conjugate boundary conditions:
\be
\Psi _2
\sim\left\{
\begin{array}{ll}
\mathcal{T}_{\rm in}^*(r-r_+)^{\ii \alpha (\omega-\varpi)} +\mathcal{T}_{\rm out}^*(r-r_+)^{-\ii \alpha (\omega-\varpi)} & {\rm as}\ r\rightarrow r_+ \,, \\
\mathcal{R}^*{\rm e}^{-\ii {\omega r}}+ {\rm e}^{\ii {\omega r}}&
{\rm as}\ r\rightarrow \infty\,.
\end{array}
\right. \label{bound3}
\ee
These two solutions are linearly independent, and the standard theory
of ODEs implies that their Wronskian $W\equiv \Psi _1
\partial_{r_*}\Psi _2 -\Psi _2 \partial_{r_*}\Psi _1$ is a constant
(independent of $r$).  If we evaluate the Wronskian near the horizon,
we get
\be
W=
2\ii(\omega-\varpi)\left (|\mathcal{T}_{\rm in}|^2-|\mathcal{T}_{\rm out}|^2\right )\,, 
\ee
and near infinity we find
\be
W=-2\ii \omega(|\mathcal{R}|^2-1)\,,
\ee
where we used $dr_*/dr=\frac{\alpha}{r-r_+}$. Equating the two yields
\be
 |\mathcal{R}|^2=1-\frac{\omega -\varpi}{\omega}\left (|\mathcal{T}_{\rm in}|^2-|\mathcal{T}_{\rm out}|^2\right )\,.
\ee
The reflection coefficient $|\mathcal{R}|^2$ is usually less than
unity, but there are some notable exceptions:
\begin{itemize}

\item If $|\mathcal{T}_{\rm in}|=|\mathcal{T}_{\rm out}|$, i.e., if
  there is no absorption by the black hole, then $|\mathcal{R}|=1$.

\item If $|\mathcal{T}_{\rm in}|>|\mathcal{T}_{\rm out}|$, i.e., if
  the hole absorbs more than what it gives away, there is
  superradiance in the regime $\omega<\varpi$. Indeed, for
  $\omega-\varpi<0$ we have that $|\mathcal{R}|^2>1$. Such a scattering
  process, where the reflected wave has actually been amplified, is
  known as superradiance. Of course the excess energy in the reflected
  wave must come from energy ``stored in the black hole'', which
  therefore decreases.

\item If $|\mathcal{T}_{\rm in}|<|\mathcal{T}_{\rm out}|$, there is
  superradiance in the other regime, but this condition also means
  that there is energy coming out of the black hole, so it is not
  surprising to see superradiance.

\end{itemize}

For the Schwarzschild spacetime $\varpi=0$, and only waves going into
the horizon are allowed ($\mathcal{T}_{\rm out}=0$). Therefore
\be
|\mathcal{R}|^2=1-\left (|\mathcal{T}_{\rm in}|^2\right )\,.
\ee
This is simply summarized by saying that energy is conserved.

If we impose the physical requirement that no waves should come out of
the horizon ($\mathcal{T}_{\rm out}=0$) and we set $\varpi=m\Omega$, as
appropriate for superradiance in the Kerr metric, we get
\be
 |\mathcal{R}|^2=1-\left(1 - \f{m\Omega}{\omega}\right)|\mathcal{T}_{\rm in}|^2\,.
\ee
Therefore superradiance in the Kerr spacetime will occur whenever
\be\label{KerrSuperradiance}
0<\omega<m\Omega\,.
\ee
The amount of superradiant amplification depends on the specific field
perturbing the black hole, and it must be obtained by direct
integration of the wave equation.

There is no superradiance for fermions: can you tell why?

\subsection{Massive Scalar Fields in the Kerr Metric}

As a prototype of the general treatment of superradiance provided
above, consider the Klein-Gordon equation (\ref{kg}) describing
massive scalar field perturbations in the Kerr metric. In
Boyer-Lindquist coordinates, the equation reads (see
e.g.~\cite{Dolan:2007mj})
\beq
&&\left[ \f{(r^2+a^2)^2}{\Delta}-a^2\sin^2\theta \right]
\f{\p^2 \Phi}{\p t^2}
+\f{4iMamr}{\Delta}
\f{\p \Phi}{\p t}\\
&&-\f{\p}{\p r}\left( \Delta \f{\p \Phi}{\p r} \right)
-\f{1}{\sin\theta}\f{\p}{\p \theta}
\left( \sin\theta \f{\p \Phi}{\p \theta} \right)
-m^2\left[\f{a^2}{\Delta}-\f{1}{\sin^2\theta}\right]\Phi
+\mu^2 \Sigma^2 \Phi
=0\,,\nn
\eeq
where $a=J/M$, $\Delta\equiv r^2-2Mr+a^2$ and we assumed an azimuthal
dependence $\Phi \sim e^{im \phi}$. The black-hole inner and outer
horizons are located at the zeros of $\Delta$, namely
$r_\pm=M\pm\sqrt{M^2-a^2}$. Focus for simplicity on the case $\mu=0$.
Assuming a harmonic time dependence and using spheroidal wave
functions \cite{flammer,Berti:2005gp} to separate the angular
dependence,
\be 
\Phi=\int d\omega e^{-i\omega t}\sum_{\ell,m} e^{im\phi} R_{\ell m}(r,\omega) 
S_{\ell m}(\theta,\omega)\,,
\ee
leads to the radial wave equation
\be
\f{d^2R_{\ell m}}{dr_*^2}+
\left[ \f{K^2+(2am\omega-a^2\omega^2-E)\Delta}{(r^2+a^2)^2}-\f{dG}{dr_*}-G^2 \right]R_{\ell m}
=0\,,
\ee
where
\beq
&&K=(r^2+a^2)\omega-am\,,\\
&&G=\f{r\Delta}{(r^2+a^2)^2}\,,\\
&&\f{dr_*}{dr}=\f{r^2+a^2}{\Delta}\,.
\eeq
$E$ is an angular separation constant, equal to $\ell(\ell+1)$ when
$a=0$ and obtained in more complicated ways when $a\neq 0$
\cite{Berti:2005gp}. 

We can easily see that this potential satisfies the boundary
conditions (\ref{bound}) for superradiance. As $r\to\infty$,
$\Delta\sim r^2$, $K\sim \omega r^2$, $G\sim r^{-1}$, $dr_*/dr\sim 1$,
$dG/dr_*\sim dG/dr\sim r^{-2}$, and therefore the potential $V\sim
K^2/(r^2+a^2)^2\sim (\omega r^2)^2/r^4\sim \omega^2$.

As $r\to r_+$, $\Delta=(r-r_+)(r-r_-)\to 0$, $(r^2+a^2)\to
2Mr_+$. Therefore $G\to 0$, $dG/dr_*\sim
\left[\Delta/(r^2+a^2)\right]dG/dr\to 0$, and the dominant term is
\be 
\left(\f{K}{r^2+a^2}\right)^2
\sim 
\left[\f{2Mr_+\omega -am}{2Mr_+}\right]^2= 
\left(\omega-m\f{a}{2Mr_+}\right)^2=
(\omega-m\Omega)^2\,.
\ee

Now, the previous treatment implies that
\begin{equation} R_{\ell m}
\sim\left\{
\begin{array}{ll}
\mathcal{T} e^{-\ii(\omega-m\Omega)r_*}
&
{\rm as}\ r\rightarrow r_+\,, \\
\mathcal{R}{\rm e}^{\ii {\omega r_*}}+ {\rm e}^{-\ii {\omega r_*}}
&
{\rm as}\ r\rightarrow \infty\,.
\end{array}
\right.
\end{equation}

\begin{figure*}[t]
\begin{center}
\epsfig{file=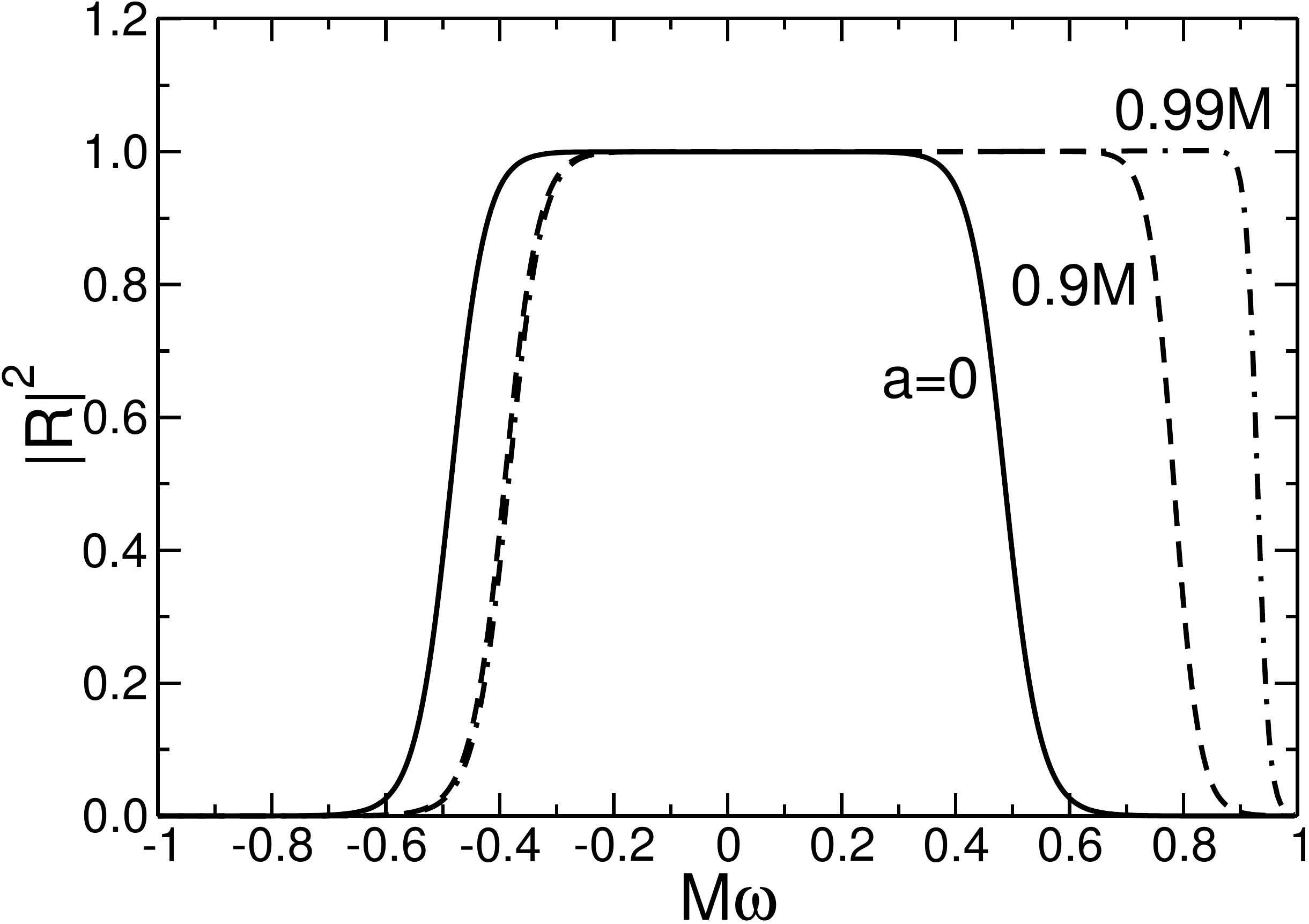,width=0.9\textwidth,angle=0,clip=true}
\epsfig{file=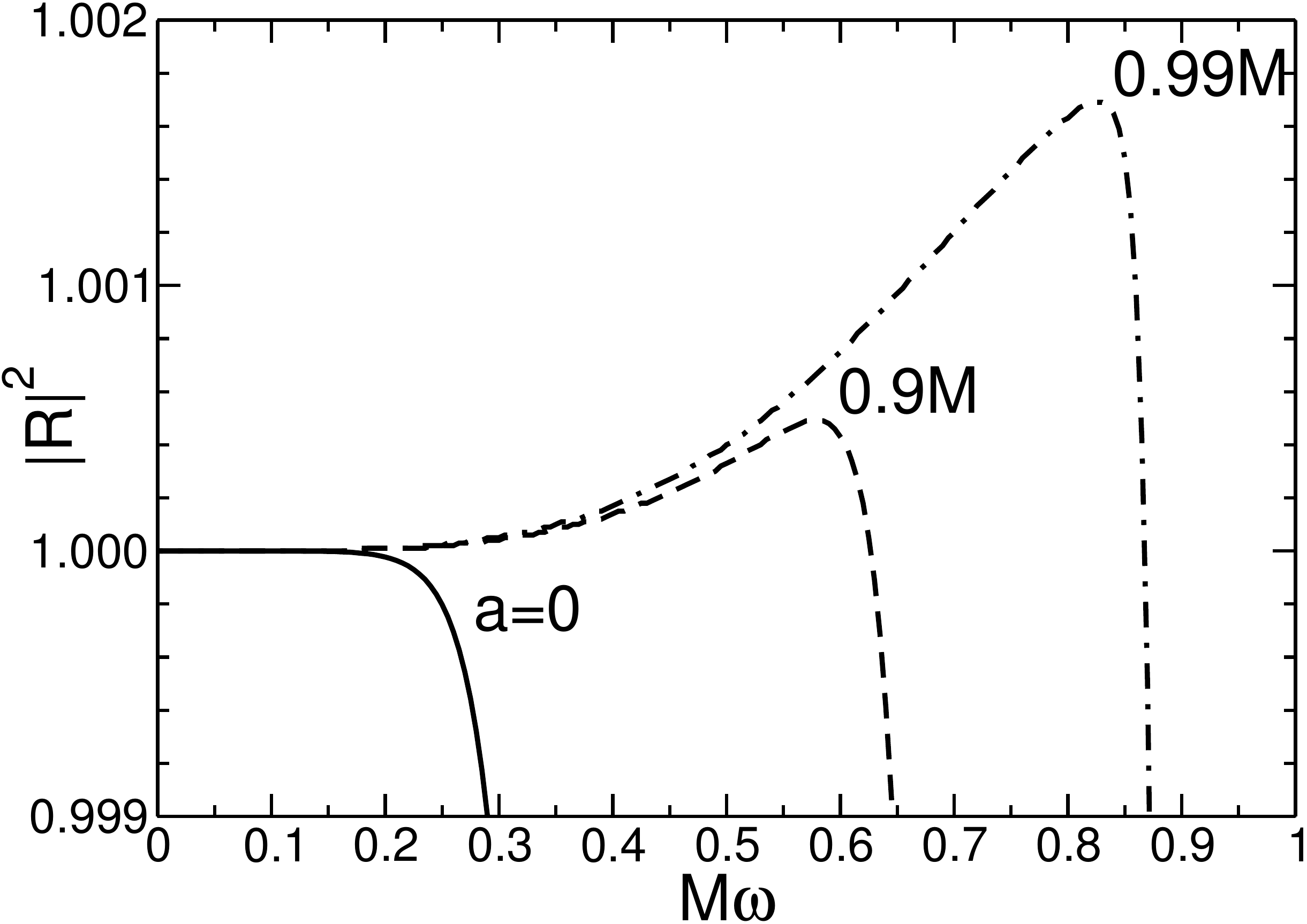,width=0.9\textwidth,angle=0,clip=true}
\caption{Superradiant amplification of scalar waves with $l=m=2$ in
  Kerr black holes (cf. Fig.~1 of
  \cite{Andersson:1998swa}). Superradiant amplification occurs when
  $M\omega\leq m(M\Omega)$. The bottom panel shows a close-up of the
  superradiant regime ($M\omega \leq 0.627$ for $a/M=0.9$, and
  $M\omega \leq 0.868$ for $a/M=0.99$).
\label{fig:superr}}
\end{center}
\end{figure*}

The superradiant amplification for scalar perturbations with $l=m=2$
is shown in Fig.~\ref{fig:superr} (where I reproduce the results
obtained in \cite{Andersson:1998swa} by an independent
frequency-domain numerical code). As a general rule, superradiant
instabilities get stronger as the spin of the perturbing field
increases.

%

%
\chapter{The Unreasonable Power of Perturbation Theory: Two Examples}
\label{ch:applications}
\allowdisplaybreaks
\minitoc

\section{Critical Phenomena in Binary Mergers}

The theory of Lyapunov exponents (Section \ref{sec:lyapunov}) has
fascinating applications in the study of critical phenomena in
black-hole binaries. In the encounter of two black holes, three
outcomes are possible: scattering for large values of the impact
parameter of the collision; merger for small values of the impact
parameter; and a {\em delayed merger} in an intermediate regime, where
the black holes can revolve around each other (in principle) an
infinite number of times $N$ by fine-tuning the impact parameter
around some critical value $b=b_*$.

I will illustrate this possibility by first considering the simple
case of extreme mass-ratio binaries (i.e., to a first level of
approximation, point particles moving along geodesics in a black-hole
spacetime), and then by reporting results from numerical simulations
of comparable-mass black hole binary encounters.

\subsection{Extreme Mass-Ratio Binaries}

Consider equatorial ($\theta=\pi/2$) timelike geodesic in the
Schwarzschild background. From Eqs.~(\ref{tdot}), (\ref{phidot}) and
(\ref{Veff}) with $\delta_1=1$ we have
\beq
\dot{t}&=&\f{E}{f}\,,\\
\dot{\phi}&=&\f{L}{r^2}\,,\\
\dot{r}^2&=&E^2-f\left(1+\frac{L^2}{r^2}\right)\equiv E^2-V_{\rm part}(r)\,.
\eeq
where $f(r)=1-2M/r$, we used the definition (\ref{effpot}) for
timelike particles, and (as usual) dots stand for derivatives with
respect to proper time $\tau$. Recall that $E$ is the particle's
energy at infinity {\em per unit mass} $\mu$, so $E=1$ corresponds to
an infall from rest and the limit $E\gg 1$ corresponds to
ultrarelativistic motion.

Geodesics can be classified according to how their energy compares to
the maximum value of the effective potential (\ref{effpot}). A simple
calculation shows that the maximum is located at
\be
r=\f{L^2-\sqrt{L^4-12L^2M^2}}{2M}\,,
\ee
and that the potential at the maximum has the value
\be V^{\rm max}_{\rm part}(L)=
\frac{1}{54}\left[\frac{L^2}{M^2}+36+\left(\frac{L^2}{M^2}-12\right)
\sqrt{1-\frac{12M^2}{L^2}}\right]. \label{max-Veff} \ee
The scattering threshold is then defined by the condition $E^2=V^{\rm
  max}_{\rm part}$: orbits with $E^2>V^{\rm max}_{\rm part}$ are
captured, while those with $E^2<V^{\rm max}_{\rm part}$ are
scattered. Given some $E$, the critical radius or impact parameter
$b_{\rm crit}$ that defines the scattering threshold is obtained by
solving the scattering threshold condition $E^2=V^{\rm max}_{\rm
  part}(L_{\rm crit})$ (in general, numerically) to obtain $L_{\rm
  crit}(E)$, and then using the following definition of the impact
parameter $b$
\be 
b = \f{L}{\sqrt{E^{2} - 1}}\,, \label{b-def}
\ee
to obtain $b_{\rm crit}$.

\begin{figure}[htb]
\includegraphics[width=\textwidth,clip=true]{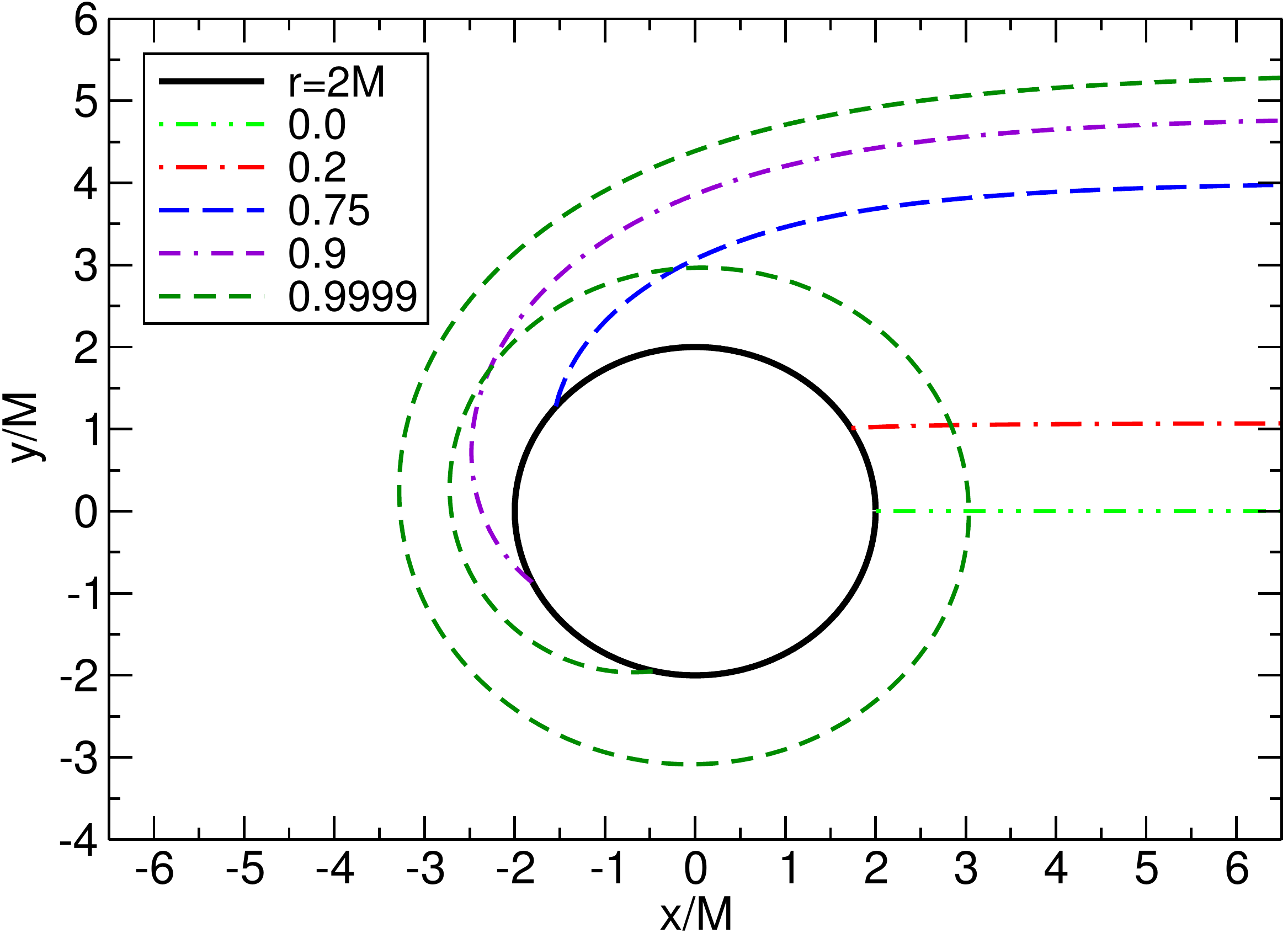}
\caption{\label{traj} Trajectories for different values of $L/L_{\rm crit}$
  (as indicated in the legend) and $E=3$. The black circle of radius $2$ marks
  the location of the horizon. (From Ref.~\cite{Berti:2010ce}.)}
\end{figure}

The geodesic equations listed above can be integrated numerically for
chosen values of $E$ and $L$.  For $L>L_{\rm crit}$ the particle does
not plunge, but rather scatters to infinity.  Examples of plunging
orbits for $E=3$ and different angular momenta $L<L_{\rm crit}$ (i.e.,
different impact parameters) are shown in Fig.~\ref{traj} in
Cartesian-like coordinates $x=r\cos\phi$, $y=r\sin\phi$. The number of
revolutions around the black hole before plunge increases as $L\to
L_{\rm crit}$.
\begin{figure}[htb]
\begin{center}
\includegraphics[width=\textwidth,clip=true]{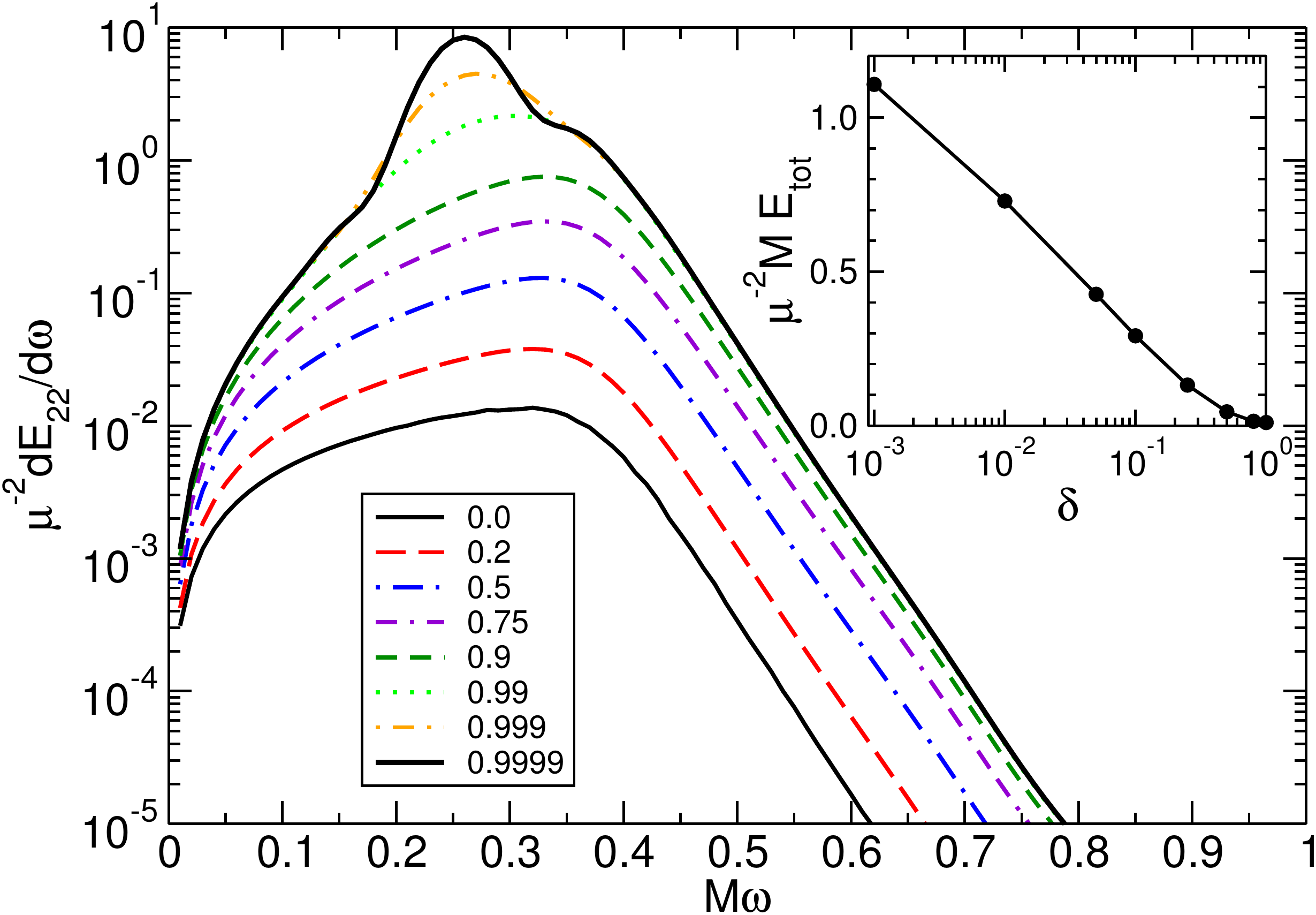}
\end{center}
\caption{Infall from rest: spectra for $l=m=2$ as $L/L_{\rm crit}\to
  1$. In the inset: logarithmic divergence of the total radiated
  energy in the same limit. (From
  Ref.~\cite{Berti:2010ce}.) \label{fig:bump}}
\end{figure}

Using black-hole perturbation theory, it is possible to compute the
energy radiated by particles falling into the black hole with
different values of $(E,\,L)$. The details of the calculation can be
found in \cite{Berti:2010ce} and references therein. In the frequency
domain, the radiated energy can be decomposed as a sum
\be
E=\sum_{l=2}^\infty \sum_{m=-l}^l \int_0^\infty d\omega \f{dE_{lm}}{d\omega}\,,
\ee
where the indices $(l,\,m)$ refer to the angular functions used to
separate variables (here spin-weighted spherical harmonics of spin
weight two). The dominant contribution to the radiation comes from
modes with $l=m=2$. Figure~\ref{fig:bump} shows the energy spectrum of
the $l=m=2$ mode for a particle with $E=1$ (infall from rest) and
different values of the orbital angular momentum $L/L_{\rm crit}$.
For $L\ll L_{\rm crit}$ the energy spectra show a simple structure,
first identified for radial infalls ($L=0$) in \cite{Davis:1971gg}:
the spectrum vanishes at low frequencies, it has a peak, and then an
exponential decay at frequencies larger than the black hole's
fundamental quasinormal mode frequency for the given $l$. However,
additional features appear when $L$ grows, and the nature of the
spectra changes quite significantly as $L/L_{\rm crit}\to 1$. As shown
in Fig.~\ref{fig:bump}, as we fine-tune the angular momentum to the
critical value separating plunge and scattering the energy spectra
with $L/L_{\rm crit}=0.99,\,\dots,\,0.9999$ clearly display a second
peak, which is related to the orbital motion of the particle. This
very distinctive ``bump'' appears at a frequency slightly {\it lower}
than the quasinormal mode frequency, significantly enhancing the
radiated energy. The location of the ``bump'' corresponds to (twice)
the orbital frequency of the particle at the marginally bound orbit,
i.e.  $\omega=2\Omega_{\rm mb}=(4M)^{-1}$. The reason for this local
maximum in the spectrum is that when $L \to L_{\rm crit}$ the particle
orbits a large number of times close to the circular marginally bound
orbit with radius $r=r_{\rm mb}$ (here $r=4M$ because $E=1$),
eventually taking an infinite amount of proper time to reach the
horizon.  The proximity of the orbit to criticality is conveniently
described by a small dimensionless ``criticality parameter''
\be
\delta\equiv 1-\frac{L}{L_{\rm crit}}\,. 
\label{deltaLcrit}
\ee
In the limit $\delta\to 0$, how many times does the particle hover at
the marginally bound circular geodesics before plunging? The answer is
given by the Lyapunov exponent calculation of Chapter
\ref{ch:particles} (see e.g.~\cite{Pretorius:2007jn}). To linear
order, the growth of the perturbation around the marginally bound
orbit is described by
\be\label{temp1}
\delta r(t)=\delta r_{\rm mb} e^{\lambda_0 t}\,,
\ee
so $\ln|\delta r(t)|=\ln|\delta r_{\rm mb}|+\lambda_0 t$. Perturbation
theory breaks down when $\delta r(t)\approx 1$, and by that time the
number of orbits that have been completed is $N\simeq \Omega_{\rm
  mb}t/(2\pi)$. Furthermore, $\delta r_{\rm mb}=k\delta$ (with $k$
some constant) for geodesics approaching the marginally bound
orbit. Taking the natural logarithm of (\ref{temp1}) and substituting
these relations yields
\be
0
=\ln|\delta r_{\rm mb}|+\lambda_0 \f{2\pi N}{\Omega_{\rm mb}}
=\ln|k\delta|+\lambda_0 \f{2\pi N}{\Omega_{\rm mb}}
\,.
\ee
Now recall our result (\ref{LyapunovGeneral}) for the Lyapunov
exponent to get:
\be
0
=\ln|k\delta|+\sqrt{\frac{V_{\rm part}''}{2 \dot{t}_{\rm mb}^2}} \f{2\pi N}{\Omega_{\rm mb}}\,.
\ee
Therefore the particle orbits the black hole
\be 
N\simeq 
-\f{\Omega_{\rm mb}\dot{t}_{\rm mb}\, \ln{|k\delta|}}
{\pi \sqrt{2V_{\rm part}''}}
\,
\ee
times before plunging \cite{Berti:2009bk} (recall that dots stand for
derivatives with respect to proper time). This logarithmic scaling
with the ``criticality parameter'' $\delta$ is typical of many
phenomena in physics (see e.g.  \cite{Choptuik:1992jv} for the
discovery that critical phenomena occur in gravitational collapse, and
\cite{Gundlach:2007gc} for a review of subsequent work in the field).

For the Schwarzschild effective potential, $r^5V_{\rm
  part}''=24ML^2-6rL^2+4Mr^2$,
and the angular velocity $\Omega_{\rm mb}$ is nothing but
$d\phi/dt$ evaluated at the marginally bound orbit
$r=r_{\rm mb}$.
When $E=1$ the marginally bound orbit is located at $r_{\rm mb}=4M$,
$M\Omega_{\rm mb}=8^{-1}$ and 
\be\label{NE1}
N\sim -\frac{1}{\pi\sqrt{2}}\ln|\delta|\,.
\ee
The inset of Fig.~\ref{fig:bump} shows that the {\it total} radiated
energy in the limit $L\to L_{\rm crit}$ does indeed scale
logarithmically with $\delta$, and hence linearly with the number of
orbits $N$, as expected from Eq.~(\ref{NE1}) for radiation from a
particle in circular orbit at the marginally bound geodesic.

In the ultrarelativistic limit $E\to \infty$ the marginally bound orbit is
located at the light ring $r_{\rm mb}=3M$, the corresponding orbital frequency
$M\Omega_{\rm mb}=(3\sqrt{3})^{-1}$ and
\be 
N\sim -\frac{1}{2\pi}\ln\left|2\delta\right|\,.
\ee
As argued in Chapter \ref{ch:waves}, the orbital frequency at the
light ring is intimately related with the eikonal (long-wavelength)
approximation of the fundamental quasinormal mode frequency of a black
hole (see also
\cite{pressringdown,mashhoon,Berti:2005eb,Cardoso:2008bp}). This
implies that ultrarelativistic infalls with near-critical impact
parameter are in a sense the most ``natural'' and efficient process to
resonantly excite the dynamics of a black hole. The proper oscillation
modes of a Schwarzschild black hole cannot be excited by particles on
{\it stable} circular orbits (i.e. particles with orbital radii $r>6M$
in Schwarzschild coordinates), but near-critical ultrarelativistic
infalls are such that the orbital ``bump'' visible in Figure
\ref{fig:bump} moves just slightly to the right to overlap with the
``knee'' due to quasinormal ringing. So ultrarelativistic infalls have
just the right orbital frequency to excite black hole oscillations.

\subsection{Comparable-Mass Binaries}

Pretorius and Khurana \cite{Pretorius:2007jn} demonstrated that
critical behavior of the kind shown above, where the number of orbits
scales logarithmically with some criticality parameter $\delta$,
occurs also in the encounters of comparable-mass black hole mergers.
Now the question is: suppose that two comparable-mass black holes
collide close to the speed of light, so that all the center-of-mass
energy of the system is kinetic energy. Is it possible to radiate all
of the energy of the system by fine-tuning the impact parameter near
threshold? And if so, what is the final state of the collision?
As demonstrated by numerical relativity simulations in a series of
papers \cite{Sperhake:2008ga,Sperhake:2009jz,Sperhake:2012me}, the
answer to this question is ``no'', and the reason is that the black
holes absorb a significant fraction of the energy during a close
encounter.

\begin{figure*}[tb]
\begin{center}
\epsfig{file=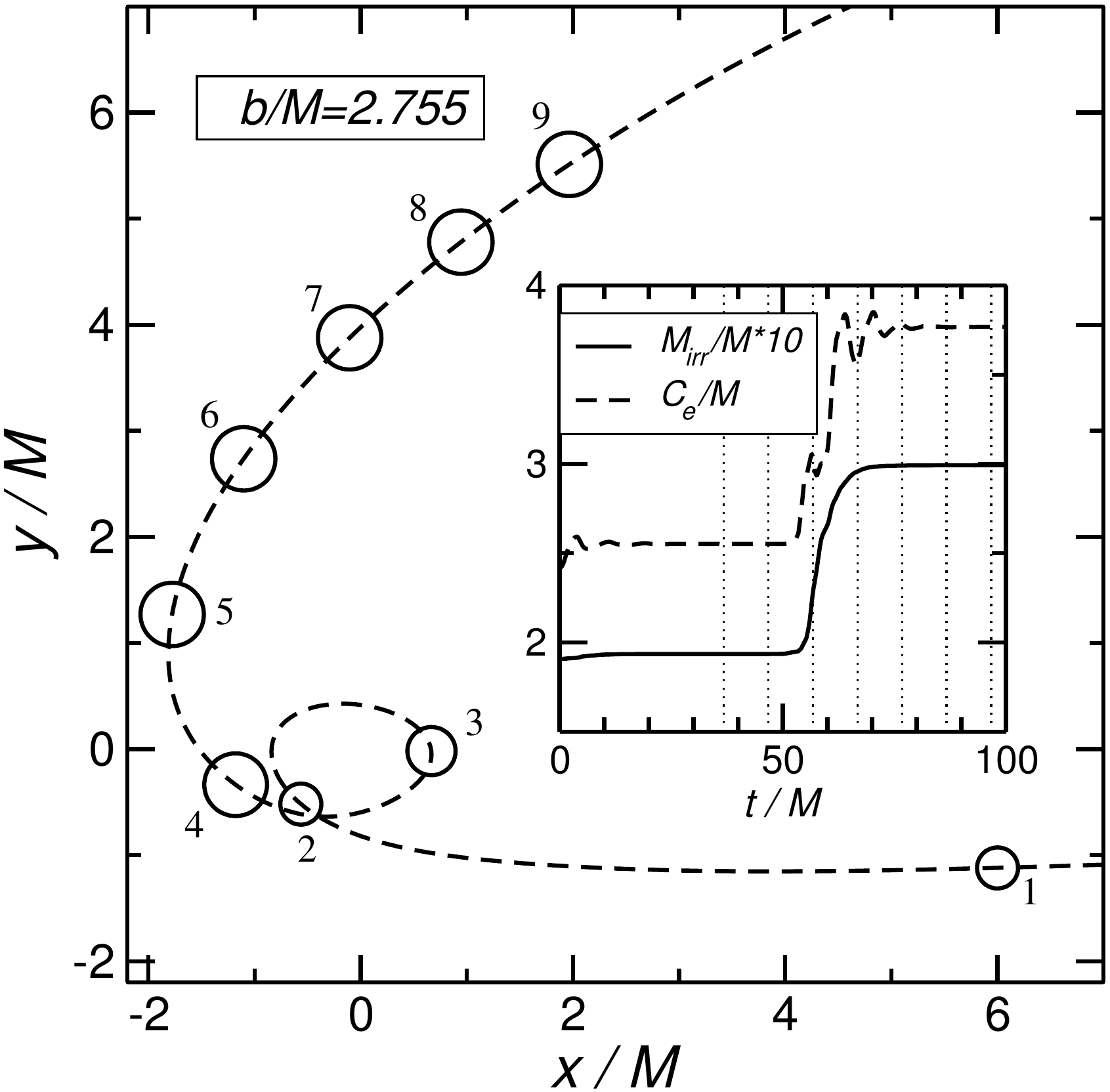,width=0.9\textwidth,clip=true} 
\caption{Trajectory of one black hole from numerical simulations of
  equal-mass ultrarelativistic black-hole encounters with a
  ``near-critical'' impact parameter $b/M=2.755$. Inset: time
  evolution of the irreducible mass $M_{\rm irr}$ and of the
  circumferential radius $C_{\rm e}$ of each hole. The circles
  represent the black hole location at intervals $\Delta t=10~M$
  (corresponding to vertical lines in the inset) and have radius equal
  to $M_{\rm irr}$. (From Ref.~\cite{Sperhake:2012me}.)}
\label{fig:triple}
\end{center}
\end{figure*}

The encounter of two equal-mass black holes is illustrated in
Fig.~\ref{fig:triple} (from \cite{Sperhake:2012me}). The plot shows
the trajectory of one of the two black holes (the trajectory of the
other hole is reflection-symmetric with respect to that of the first
hole).  In the numerical simulations we monitor the apparent horizon
dynamics of the individual holes by measuring the equatorial
circumference $C_{\rm e}=4\pi M$ and the irreducible mass $M_{\rm
  irr}$ of each black hole before and after the encounter. The inset
of the right panel of Fig.~\ref{fig:triple} shows the variation of
these quantities with time: absorption occurs over a short timescale
$\approx 10M$.
Since the apparent horizon area $A_{\rm AH}=16\pi M_{\rm
  irr}^2=[C_{\rm e}^2/(2\pi)](1+\sqrt{1-\chi^2})$, in this way we can
estimate the rest mass and dimensionless spin $\chi=a/M$ of each hole
before ($M_{\rm i}$, $\chi_{\rm i}$) and after ($M_{\rm s}$,
$\chi_{\rm s}$) the first encounter. We define the absorbed energy
$E_{\rm abs}=2(M_s-M_i)$.  It turns out that $(E_{\rm rad}+E_{\rm
  abs})/M$ accounts for most of the total available kinetic energy in
the system, and therefore the system is no longer kinetic-energy
dominated after the encounter. A fit of the data yields $E_{\rm
  rad}/K=0.46 (1 + 1.4/\gamma^2)$ and $E_{\rm
  abs}/K=0.55(1-1/\gamma)$, where $\gamma$ is the Lorentz boost
parameter, suggesting that radiation and absorption contribute about
equally in the ultrarelativistic limit, and therefore that absorption
sets an upper bound $\approx 50\%$ on the maximum energy that can be
radiated.

The fact that absorption and emission are comparable in the
ultrarelativistic limit is supported by point-particle calculations in
black hole perturbation theory. For example, Misner et
al. \cite{Misner:1972jf} studied the radiation from ultrarelativistic
particles in circular orbits near the Schwarzschild light ring,
i.e. at $r=3M(1+\epsilon)$. Using a scalar-field model they found that
50\% of the radiation is absorbed and 50\% is radiated as $\epsilon\to
0$. The same conclusion applies to {\em gravitational} perturbations
of Schwarzschild black holes when one ignores radiation reaction
(self-force) effects, and a recent analysis including self-force
effects finds that $42\%$ of the energy should be absorbed by
nonrotating black holes as $\epsilon\to 0$ (cf. Fig.~4 of
\cite{Gundlach:2012aj}).

\clearpage

\section{Black-Hole Bombs}
\label{sec:BHBombs}

The superradiant amplification mechanism discussed in Section
\ref{subs:superradiance} has a rather dramatic consequence for
astrophysics and particle physics: measurements of the spin of massive
black holes, the largest and simplest macroscopic objects in the
Universe, can be used to set upper bounds on the mass of light bosonic
particles. The reason behind this striking connection between the
smallest and largest objects in our physical world is the so-called
``black-hole bomb'' instability first investigated by Press and
Teukolsky \cite{Press:1972zz} (see also \cite{Cardoso:2004nk}, and
\cite{Shlapentokh-Rothman:2013ysa} for a recent rigorous proof for
scalar perturbations).

The idea is the following. Imagine surrounding a rotating black hole
by a perfectly reflecting mirror. An ingoing monochromatic wave with
frequency in the superradiant regime defined by
Eq.~(\ref{KerrSuperradiance}) will be reflected and amplified at the
expense of the rotational energy of the black hole, then it will be
reflected again by the mirror and amplified once more. The process
keeps repeating and triggers a runaway growth of the perturbation -- a
``black-hole bomb''. The end state cannot be predicted in linearized
perturbation theory, but it is reasonable to expect that we will be
left with a slowly rotating (or nonrotating) black hole, transferring
the whole rotational energy of the hole (a huge amount compared to
nuclear physics standards!) to the field.

The original Press-Teukolsky mechanism is of mostly speculative
interest: for example, one could imagine a very advanced civilization
building perfectly reflective mirrors around rotating black holes to
solve their energy problems. Luckily, Nature gives us actual
``mirrors'' in the form of massive bosonic fields. As shown in
Fig.~\ref{fig:SchwPot} for massive scalars, whenever the field has
mass the effective potential for wave propagation tends to a nonzero
value at infinity. This potential barrier is the particle-physics
analog of Press and Teukolsky's perfectly reflecting mirror.

This simple consideration has immediate implications for fundamental
physics. Many proposed extensions of the Standard Model predicted the
existence of ultralight bosons, such as the light scalars with
$10^{-33}~{\rm eV}<m<10^{-18}~{\rm eV}$ of the ``string axiverse''
scenario \cite{Arvanitaki:2010sy}, ``hidden photons''
\cite{Goldhaber:2008xy,Goodsell:2009xc,Jaeckel:2010ni,Camara:2011jg}
and massive gravitons
\cite{Goldhaber:2008xy,Hinterbichler:2011tt,deRham:2014zqa}. Explicit
calculations of the instability timescales have been performed for
massive scalar
\cite{Press:1972zz,Damour:1976kh,Zouros:1979iw,Detweiler:1980uk,Dolan:2007mj,Rosa:2009ei,Cardoso:2011xi},
vector (Proca) \cite{Pani:2012vp,Pani:2012bp} and tensor (massive
graviton) \cite{Brito:2013wya} perturbations of a Kerr black hole. In
all cases the instability is regulated by the dimensionless parameter
$M\mu$ (in units $G=c=1$), where $M$ is the BH mass and $m=\mu\hbar$
is the bosonic field mass, and it is strongest for maximally spinning
BHs when $M\mu\sim 1$.  As discussed around Eq.~(\ref{lambdaC}), for a
solar-mass black hole and a field of mass $m\sim 1$~eV the parameter
$M\mu\sim 10^{10}$. In this case the instability is exponentially
suppressed~\cite{Zouros:1979iw}, and the instability timescale would
be larger than the age of the Universe. The strongest superradiant
instabilities develop when $M\mu\sim 1$, i.e. when the Compton
wavelength of the perturbing field is comparable to the ``size'' of
the black hole's event horizon. This can occur for light primordial
BHs that may have been produced in the early Universe, or for the
ultralight exotic particles proposed in some extensions of the
Standard Model. In particular, fields of mass $m\sim 10^{-20}$~eV
around the heaviest supermassive black holes ($M\sim 10^{10}M_\odot$)
are ideal candidates.

\begin{figure}[bht]
\begin{center}
\epsfig{file=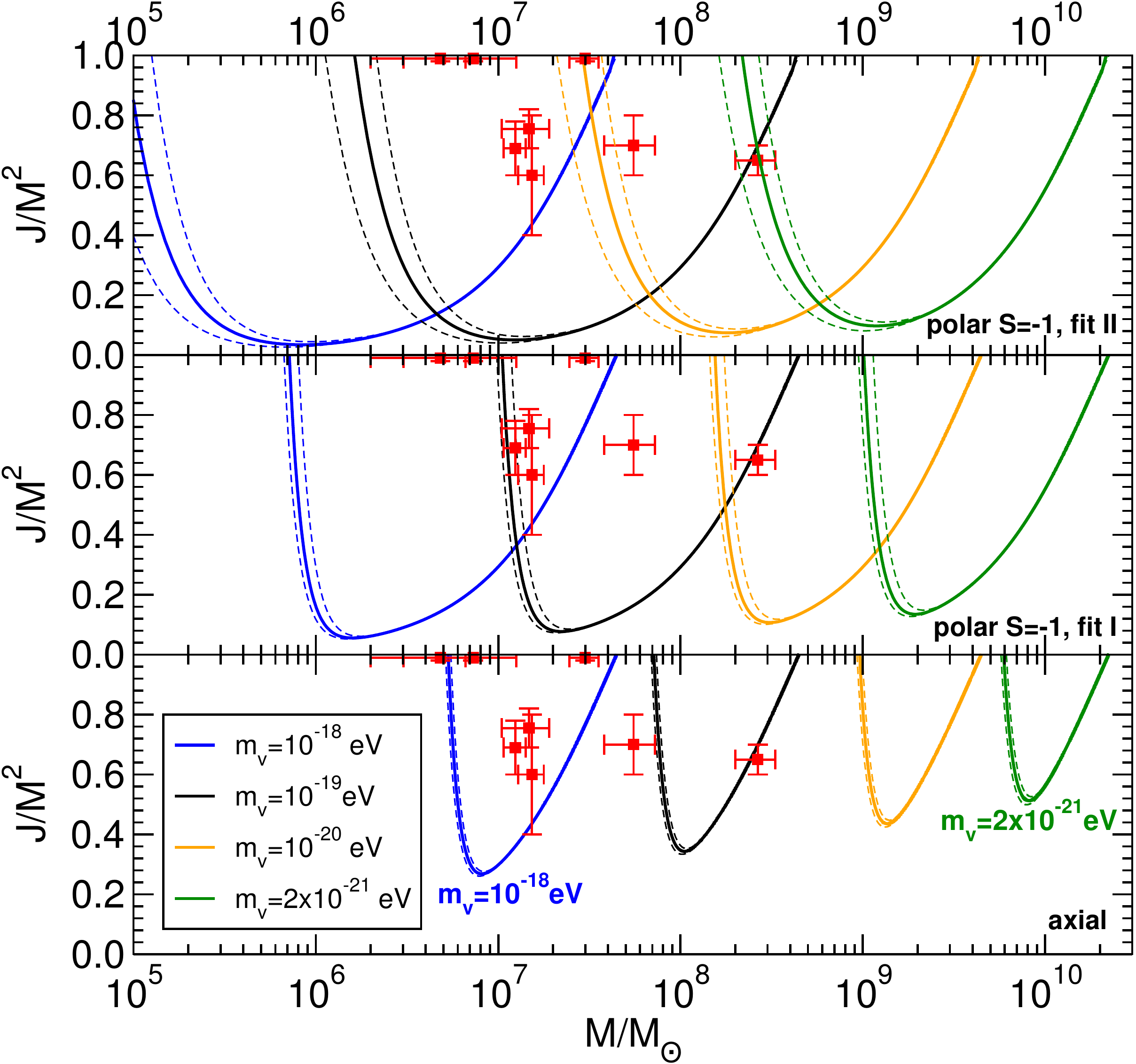,width=0.85\textwidth,clip=true}
\caption{(color online) Contour plots in the ``BH Regge
  plane''~\cite{Arvanitaki:2010sy} corresponding to an instability
  timescale shorter than a typical accretion timescale, $\tau_{\rm
    Salpeter}=4.5\times 10^7$~yr, for different values of the vector
  (Proca) field mass $m_v={{\mu}}\hbar$ (from left to right:
  $m_v=10^{-18}{\rm eV}$, $10^{-19}{\rm eV}$, $10^{-20}{\rm eV}$,
  $2\times10^{-21}{\rm eV}$). The instability operates for both polar
  (even parity) and axial (odd parity) modes. For polar modes in the
  $S=-1$ polarization, which provides the strongest instability, we
  use two different fits (top and middle panel) to numerical
  calculations of the instability (see \cite{Pani:2012vp,Pani:2012bp}
  for details); axial mode instability windows are shown in the bottom
  panel.  Dashed lines bracket our estimated numerical errors.  The
  experimental points (with error bars) refer to the mass and spin
  estimates of supermassive BHs listed in Table~2
  of~\cite{Brenneman:2011wz}. The rightmost point corresponds to the
  supermassive BH in Fairall~9~\cite{Schmoll:2009gq}. Supermassive BHs
  lying above each of these curves would be unstable on an observable
  timescale, and therefore they exclude a whole range of Proca field
  masses.\label{fig:ReggePlane} }
\end{center}
\end{figure}

This yields stringent constraints on the mass $m$ of the perturbing
field. If there were fields of mass $m$ around a supermassive rotating
black hole, the superradiant instability would reduce their spin on a
timescale that can be computed using black-hole perturbation theory
(this timescale is shortest when $M\mu\sim 1$). We can exclude the
existence of a field of mass $m$ whenever the superradiant instability
timescale is smaller than the Salpeter time, i.e. the typical time
over which accretion could potentially spin up the hole
\cite{Salpeter:1964kb}. By setting the Salpeter timescale equal to the
instability timescale we can draw ``instability windows'' such as
those shown in Fig.~\ref{fig:ReggePlane}; these are sometimes called
gaps in the mass-spin black-hole ``Regge
spectrum''~\cite{Arvanitaki:2010sy}. Any measurement of a black-hole
spin with value above one of the instability windows excludes a whole
range of masses for the perturbing field.

It is worth mentioning that massive fields can have other potentially
observable signatures. They can modify the inspiral dynamics of
compact binaries~\cite{Alsing:2011er} and even freeze the inspiral of
compact objects into massive black holes, creating ``floating orbits''
that produce stimulated emission of radiation by extracting the hole's
rotational energy (the gravitational-wave equivalent of a
laser!)~\cite{Cardoso:2011xi,Yunes:2011aa}. Another potentially
observable event is a ``bosenova'', i.e. a collapse of the axion cloud
producing a relatively large emission of scalar radiation (see
e.g. ~\cite{Kodama:2011zc,Yoshino:2012kn,Mocanu:2012fd,Witek:2012tr}).


\begin{thebibliography}{10}

\bibitem{Abramowitz:1970as}
M.~Abramowitz and I.~A. Stegun.
\newblock {\em Handbook of Mathematical Functions with Formulas, Graphs, and
  Mathematical Tables}.
\newblock Dover, New York, 1972.

\bibitem{Alsing:2011er}
J.~Alsing, E.~Berti, C.~M. Will, and H.~Zaglauer.
\newblock {Gravitational radiation from compact binary systems in the massive
  Brans-Dicke theory of gravity}.
\newblock {\em Phys.Rev.}, D85:064041, 2012.

\bibitem{Andersson:2003fh}
N.~Andersson and C.~J. Howls.
\newblock {The asymptotic quasinormal mode spectrum of non-rotating black
  holes}.
\newblock {\em Class. Quant. Grav.}, 21:1623--1642, 2004.

\bibitem{Andersson:1998swa}
N.~Andersson, P.~Laguna, and P.~Papadopoulos.
\newblock {Dynamics of scalar fields in the background of rotating black holes.
  2. A Note on superradiance}.
\newblock {\em Phys.Rev.}, D58:087503, 1998.

\bibitem{Arvanitaki:2010sy}
A.~Arvanitaki and S.~Dubovsky.
\newblock {Exploring the String Axiverse with Precision Black Hole Physics}.
\newblock {\em Phys.Rev.}, D83:044026, 2011.

\bibitem{Bardeen:1972fi}
J.~M. Bardeen, W.~H. Press, and S.~A. Teukolsky.
\newblock {Rotating black holes: Locally nonrotating frames, energy extraction,
  and scalar synchrotron radiation}.
\newblock {\em Astrophys. J.}, 178:347, 1972.

\bibitem{benderorszag}
C.~M. Bender and S.~A. Orszag.
\newblock {\em Advanced Mathematical Methods for Scientists and Engineers}.
\newblock McGraw-Hill, New York, 1978.

\bibitem{Berti:2004md}
E.~Berti.
\newblock {Black hole quasinormal modes: Hints of quantum gravity?}
\newblock 2004.
\newblock gr-qc/0411025.

\bibitem{Berti:2005gp}
E.~Berti, V.~Cardoso, and M.~Casals.
\newblock {Eigenvalues and eigenfunctions of spin-weighted spheroidal harmonics
  in four and higher dimensions}.
\newblock {\em Phys. Rev.}, D73:024013, 2006.

\bibitem{Berti:2009bk}
E.~Berti, V.~Cardoso, L.~Gualtieri, F.~Pretorius, and U.~Sperhake.
\newblock {Comment on 'Kerr Black Holes as Particle Accelerators to Arbitrarily
  High Energy'}.
\newblock {\em Phys. Rev. Lett.}, 103:239001, 2009.

\bibitem{Berti:2010ce}
E.~Berti, V.~Cardoso, T.~Hinderer, M.~Lemos, F.~Pretorius, et~al.
\newblock {Semianalytical estimates of scattering thresholds and gravitational
  radiation in ultrarelativistic black hole encounters}.
\newblock {\em Phys.Rev.}, D81:104048, 2010.

\bibitem{Berti:2009kk}
E.~Berti, V.~Cardoso, and A.~O. Starinets.
\newblock {Quasinormal modes of black holes and black branes}.
\newblock {\em Class.Quant.Grav.}, 26:163001, 2009.

\bibitem{Berti:2003zu}
E.~Berti and K.~D. Kokkotas.
\newblock {Asymptotic quasinormal modes of Reissner-Nordstroem and Kerr black
  holes}.
\newblock {\em Phys. Rev.}, D68:044027, 2003.

\bibitem{Berti:2005eb}
E.~Berti and K.~D. Kokkotas.
\newblock {Quasinormal modes of Kerr-Newman black holes: Coupling of
  electromagnetic and gravitational perturbations}.
\newblock {\em Phys. Rev.}, D71:124008, 2005.

\bibitem{Brenneman:2011wz}
L.~Brenneman, C.~Reynolds, M.~Nowak, R.~Reis, M.~Trippe, et~al.
\newblock {The Spin of the Supermassive Black Hole in NGC 3783}.
\newblock {\em Astrophys.J.}, 736:103, 2011.

\bibitem{Brito:2013wya}
R.~Brito, V.~Cardoso, and P.~Pani.
\newblock {Massive spin-2 fields on black hole spacetimes: Instability of the
  Schwarzschild and Kerr solutions and bounds on the graviton mass}.
\newblock {\em Phys.Rev.}, D88(2):023514, 2013.

\bibitem{Camara:2011jg}
P.~G. Camara, L.~E. Ibanez, and F.~Marchesano.
\newblock {RR photons}.
\newblock {\em JHEP}, 1109:110, 2011.

\bibitem{Cardoso:2006bv}
V.~Cardoso and M.~Cavaglia.
\newblock {Stability of naked singularities and algebraically special modes}.
\newblock {\em Phys. Rev.}, D74:024027, 2006.

\bibitem{Cardoso:2011xi}
V.~Cardoso, S.~Chakrabarti, P.~Pani, E.~Berti, and L.~Gualtieri.
\newblock {Floating and sinking: The Imprint of massive scalars around rotating
  black holes}.
\newblock {\em Phys.Rev.Lett.}, 107:241101, 2011.

\bibitem{Cardoso:2004nk}
V.~Cardoso, O.~J.~C. Dias, J.~P.~S. Lemos, and S.~Yoshida.
\newblock {The black hole bomb and superradiant instabilities}.
\newblock {\em Phys. Rev.}, D70:044039, 2004.

\bibitem{Cardoso:2008bp}
V.~Cardoso, A.~S. Miranda, E.~Berti, H.~Witek, and V.~T. Zanchin.
\newblock {Geodesic stability, Lyapunov exponents and quasinormal modes}.
\newblock {\em Phys. Rev.}, D79:064016, 2009.

\bibitem{Carroll}
S.~M. {Carroll}.
\newblock {\em {Spacetime and geometry. An introduction to general
  relativity}}.
\newblock Addison Wesley, San Francisco, CA, USA, 2004.

\bibitem{MTB}
S.~Chandrasekhar.
\newblock {\em The Mathematical Theory of Black Holes}.
\newblock Oxford University Press, New York, 1983.

\bibitem{chandraspecial}
S.~Chandrasekhar.
\newblock {On algebraically special perturbations of black hole}.
\newblock {\em Proc. R. Soc. Lond.}, A392:1, 1984.

\bibitem{RevModPhys.70.1545}
E.~S.~C. Ching, P.~T. Leung, A.~Maassen van~den Brink, W.~M. Suen, S.~S. Tong,
  and K.~Young.
\newblock Quasinormal-mode expansion for waves in open systems.
\newblock {\em Rev. Mod. Phys.}, 70(4):1545--1554, Oct 1998.

\bibitem{Choptuik:1992jv}
M.~W. Choptuik.
\newblock {Universality and scaling in gravitational collapse of a massless
  scalar field}.
\newblock {\em Phys. Rev. Lett.}, 70:9--12, 1993.

\bibitem{Cornish:2003ig}
N.~J. Cornish and J.~J. Levin.
\newblock {Lyapunov timescales and black hole binaries}.
\newblock {\em Class.Quant.Grav.}, 20:1649--1660, 2003.

\bibitem{Damour:1976kh}
T.~Damour, N.~Deruelle, and R.~Ruffini.
\newblock {On Quantum Resonances in Stationary Geometries}.
\newblock {\em Lett.Nuovo Cim.}, 15:257--262, 1976.

\bibitem{Davis:1971gg}
M.~Davis, R.~Ruffini, W.~H. Press, and R.~H. Price.
\newblock {Gravitational radiation from a particle falling radially into a
  schwarzschild black hole}.
\newblock {\em Phys. Rev. Lett.}, 27:1466--1469, 1971.

\bibitem{deRham:2014zqa}
C.~de~Rham.
\newblock {Massive Gravity}.
\newblock {\em Living Rev.Rel.}, 17:7, 2014.

\bibitem{Decanini:2009mu}
Y.~Decanini and A.~Folacci.
\newblock {Regge poles of the Schwarzschild black hole: A WKB approach}.
\newblock {\em Phys.Rev.}, D81:024031, 2010.

\bibitem{Decanini:2002ha}
Y.~Decanini, A.~Folacci, and B.~Jensen.
\newblock {Complex angular momentum in black hole physics and the quasi-normal
  modes}.
\newblock {\em Phys. Rev.}, D67:124017, 2003.

\bibitem{Detweiler:1980uk}
S.~L. Detweiler.
\newblock {Klein-Gordon Equation and Rotating Black Holes}.
\newblock {\em Phys.Rev.}, D22:2323--2326, 1980.

\bibitem{Dolan:2007mj}
S.~R. Dolan.
\newblock {Instability of the massive Klein-Gordon field on the Kerr
  spacetime}.
\newblock {\em Phys.Rev.}, D76:084001, 2007.

\bibitem{Dolan:2009nk}
S.~R. Dolan and A.~C. Ottewill.
\newblock {On an Expansion Method for Black Hole Quasinormal Modes and Regge
  Poles}.
\newblock {\em Class.Quant.Grav.}, 26:225003, 2009.

\bibitem{Ferrari:1984zz}
V.~Ferrari and B.~Mashhoon.
\newblock {New approach to the quasinormal modes of a black hole}.
\newblock {\em Phys.Rev.}, D30:295--304, 1984.

\bibitem{flammer}
C.~Flammer.
\newblock {\em {Spheroidal wave functions}}.
\newblock 1957.

\bibitem{Frolov:1998wf}
V.~P. Frolov and I.~D. Novikov.
\newblock {\em {Black hole physics: Basic concepts and new developments}}.
\newblock Dordrecht, Netherlands: Kluwer Academic (1998) 770 p.

\bibitem{Glass:1980}
E.~N. {Glass}.
\newblock {Newtonian spherical gravitational collapse}.
\newblock {\em Journal of Physics A Mathematical General}, 13:3097--3104, Sept.
  1980.

\bibitem{goebel}
C.~J. Goebel.
\newblock {Comments on the ``vibrations'' of a black hole}.
\newblock {\em Astrophys. J.}, L172:95, 1972.

\bibitem{Goldhaber:2008xy}
A.~S. Goldhaber and M.~M. Nieto.
\newblock {Photon and Graviton Mass Limits}.
\newblock {\em Rev.Mod.Phys.}, 82:939--979, 2010.

\bibitem{Goodsell:2009xc}
M.~Goodsell, J.~Jaeckel, J.~Redondo, and A.~Ringwald.
\newblock {Naturally Light Hidden Photons in LARGE Volume String
  Compactifications}.
\newblock {\em JHEP}, 0911:027, 2009.

\bibitem{Gundlach:2012aj}
C.~Gundlach, S.~Akcay, L.~Barack, and A.~Nagar.
\newblock {Critical phenomena at the threshold of immediate merger in binary
  black hole systems: the extreme mass ratio case}.
\newblock {\em Phys.Rev.}, D86:084022, 2012.

\bibitem{Gundlach:2007gc}
C.~Gundlach and J.~M. Martin-Garcia.
\newblock {Critical phenomena in gravitational collapse}.
\newblock {\em Living Rev.Rel.}, 10:5, 2007.

\bibitem{Hartle}
J.~B. {Hartle}.
\newblock {\em {Gravity : an introduction to Einstein's general relativity}}.
\newblock Addison Wesley, San Francisco, CA, USA, 2003.

\bibitem{Hinterbichler:2011tt}
K.~Hinterbichler.
\newblock {Theoretical Aspects of Massive Gravity}.
\newblock {\em Rev.Mod.Phys.}, 84:671--710, 2012.

\bibitem{Hod:1998vk}
S.~Hod.
\newblock {Bohr's correspondence principle and the area spectrum of quantum
  black holes}.
\newblock {\em Phys. Rev. Lett.}, 81:4293, 1998.

\bibitem{Ishibashi:2003ap}
A.~Ishibashi and H.~Kodama.
\newblock {Stability of higher-dimensional Schwarzschild black holes}.
\newblock {\em Prog. Theor. Phys.}, 110:901--919, 2003.

\bibitem{Iyer:1986nq}
S.~Iyer.
\newblock {Black Hole Normal Modes: A WKB Approach. 2. Schwarzschild Black
  Holes}.
\newblock {\em Phys.Rev.}, D35:3632, 1987.

\bibitem{Iyer:1986np}
S.~Iyer and C.~M. Will.
\newblock {Black Hole Normal Modes: A WKB Approach. 1. Foundations and
  Application of a Higher Order WKB Anaysis of Potential Barrier Scattering}.
\newblock {\em Phys.Rev.}, D35:3621, 1987.

\bibitem{Jaeckel:2010ni}
J.~Jaeckel and A.~Ringwald.
\newblock {The Low-Energy Frontier of Particle Physics}.
\newblock {\em Ann.Rev.Nucl.Part.Sci.}, 60:405--437, 2010.

\bibitem{Jaffe:1934}
G.~Jaff\'e.
\newblock {\em Z.~Phys.}, 87:535, 1934.

\bibitem{Kodama:2003jz}
H.~Kodama and A.~Ishibashi.
\newblock {A Master equation for gravitational perturbations of maximally
  symmetric black holes in higher dimensions}.
\newblock {\em Prog.Theor.Phys.}, 110:701--722, 2003.

\bibitem{Kodama:2003kk}
H.~Kodama and A.~Ishibashi.
\newblock {Master equations for perturbations of generalized static black holes
  with charge in higher dimensions}.
\newblock {\em Prog.Theor.Phys.}, 111:29--73, 2004.

\bibitem{Kodama:2011zc}
H.~Kodama and H.~Yoshino.
\newblock {Axiverse and Black Hole}.
\newblock {\em Int.J.Mod.Phys.Conf.Ser.}, 7:84--115, 2012.

\bibitem{Kokkotas:1999bd}
K.~D. Kokkotas and B.~G. Schmidt.
\newblock {Quasi-normal modes of stars and black holes}.
\newblock {\em Living Rev. Rel.}, 2:2, 1999.

\bibitem{Konoplya:2011qq}
R.~Konoplya and A.~Zhidenko.
\newblock {Quasinormal modes of black holes: From astrophysics to string
  theory}.
\newblock {\em Rev.Mod.Phys.}, 83:793--836, 2011.

\bibitem{Konoplya:2003ii}
R.~A. Konoplya.
\newblock {Quasinormal behavior of the d-dimensional Schwarzschild black hole
  and higher order WKB approach}.
\newblock {\em Phys. Rev.}, D68:024018, 2003.

\bibitem{Leaver:1985ax}
E.~W. Leaver.
\newblock {An Analytic representation for the quasi normal modes of Kerr black
  holes}.
\newblock {\em Proc. Roy. Soc. Lond.}, A402:285--298, 1985.

\bibitem{leJMP}
E.~W. Leaver.
\newblock {Solutions to a generalized spheroidal wave equation: Teukolsky's
  equations in general relativity, and the two-center problem in molecular
  quantum mechanics}.
\newblock {\em J. Math. Phys.}, 27:1238, 1986.

\bibitem{Leung:1999fr}
P.~Leung, A.~Maassen van~den Brink, W.~Suen, C.~Wong, and K.~Young.
\newblock {SUSY transformations for quasinormal and total transmission modes of
  open systems}.
\newblock 1999.
\newblock math-ph/9909030.

\bibitem{Mashhoon:1985}
B.~{Mashhoon}.
\newblock In H.~Ning, editor, {\em {Proceedings of the Third Marcel Grossmann
  Meeting on Recent Developments of General Relativity}}, page 599, Amsterdam,
  1983. North-Holland.

\bibitem{mashhoon}
B.~Mashhoon.
\newblock {Stability of charged rotating black holes in the eikonal
  approximation}.
\newblock {\em Phys. Rev.}, D31:290, 1985.

\bibitem{1980AnPhy.130...99M}
B.~{Mashhoon} and M.~{Hossein Partovi}.
\newblock {On the gravitational motion of a fluid obeying an equation of
  state}.
\newblock {\em Annals of Physics}, 130:99--138, Nov. 1980.

\bibitem{Misner:1972jf}
C.~W. Misner, R.~Breuer, D.~Brill, P.~Chrzanowski, H.~Hughes, et~al.
\newblock {Gravitational synchrotron radiation in the schwarzschild geometry}.
\newblock {\em Phys.Rev.Lett.}, 28:998--1001, 1972.

\bibitem{Misner:1974qy}
C.~W. Misner, K.~S. Thorne, and J.~A. Wheeler.
\newblock {\em {Gravitation}}.
\newblock W. H. Freeman, 1973.

\bibitem{Mocanu:2012fd}
G.~Mocanu and D.~Grumiller.
\newblock {Self-organized criticality in boson clouds around black holes}.
\newblock {\em Phys.Rev.}, D85:105022, 2012.

\bibitem{Motl:2002hd}
L.~Motl.
\newblock {An analytical computation of asymptotic Schwarzschild quasinormal
  frequencies}.
\newblock {\em Adv. Theor. Math. Phys.}, 6:1135--1162, 2003.

\bibitem{Motl:2003cd}
L.~Motl and A.~Neitzke.
\newblock {Asymptotic black hole quasinormal frequencies}.
\newblock {\em Adv. Theor. Math. Phys.}, 7:307--330, 2003.

\bibitem{Nollert:1993zz}
H.-P. Nollert.
\newblock {Quasinormal modes of Schwarzschild black holes: The determination of
  quasinormal frequencies with very large imaginary parts}.
\newblock {\em Phys.Rev.}, D47:5253--5258, 1993.

\bibitem{Nollert:1999ji}
H.-P. Nollert.
\newblock {Quasinormal modes: the characteristic `sound' of black holes and
  neutron stars}.
\newblock {\em Class. Quant. Grav.}, 16:R159--R216, 1999.

\bibitem{Nollert:1998ys}
H.-P. Nollert and R.~H. Price.
\newblock {Quantifying excitations of quasinormal mode systems}.
\newblock {\em J. Math. Phys.}, 40:980--1010, 1999.

\bibitem{Pani:2012vp}
P.~Pani, V.~Cardoso, L.~Gualtieri, E.~Berti, and A.~Ishibashi.
\newblock {Black hole bombs and photon mass bounds}.
\newblock {\em Phys.Rev.Lett.}, 109:131102, 2012.

\bibitem{Pani:2012bp}
P.~Pani, V.~Cardoso, L.~Gualtieri, E.~Berti, and A.~Ishibashi.
\newblock {Perturbations of slowly rotating black holes: massive vector fields
  in the Kerr metric}.
\newblock {\em Phys.Rev.}, D86:104017, 2012.

\bibitem{PoissonGray}
E.~Poisson and C.~G. Gray.
\newblock {When action is not least for orbits in general relativity}.
\newblock {\em Am.J.Phys.}, 79:43, 2011.

\bibitem{PW}
E.~{Poisson} and C.~M. {Will}.
\newblock {\em {Gravity: Newtonian, Post-Newtonian, Relativistic}}.
\newblock Cambridge University Press, Cambridge, 2014.

\bibitem{pressringdown}
W.~H. Press.
\newblock {Long wave trains of gravitational waves from a vibrating black
  hole}.
\newblock {\em Astrophys. J.}, L170:105, 1971.

\bibitem{Press:1972zz}
W.~H. Press and S.~A. Teukolsky.
\newblock {Floating Orbits, Superradiant Scattering and the Black-hole Bomb}.
\newblock {\em Nature}, 238:211--212, 1972.

\bibitem{Pretorius:2007jn}
F.~Pretorius and D.~Khurana.
\newblock {Black hole mergers and unstable circular orbits}.
\newblock {\em Class.Quant.Grav.}, 24:S83--S108, 2007.

\bibitem{Rosa:2009ei}
J.~G. Rosa.
\newblock {The Extremal black hole bomb}.
\newblock {\em JHEP}, 1006:015, 2010.

\bibitem{Ruffini}
R.~Ruffini.
\newblock {\em Black Holes: les Astres Occlus}.
\newblock Gordon and Breach Science Publishers, New York, 1973.

\bibitem{Salpeter:1964kb}
E.~Salpeter.
\newblock {Accretion of Interstellar Matter by Massive Objects}.
\newblock {\em Astrophys.J.}, 140:796--800, 1964.

\bibitem{Schmoll:2009gq}
S.~Schmoll, J.~Miller, M.~Volonteri, E.~Cackett, C.~Reynolds, et~al.
\newblock {Constraining the Spin of the Black Hole in Fairall 9 with Suzaku}.
\newblock {\em Astrophys.J.}, 703:2171--2176, 2009.

\bibitem{Schutz}
B.~{Schutz}.
\newblock {\em {A First Course in General Relativity}}.
\newblock Cambridge University Press, Cambridge, May 2009.

\bibitem{Schutz:1985km}
B.~F. Schutz and C.~M. Will.
\newblock {Black hole normal modes: A semianalytic approach}.
\newblock {\em Astrophys. J.}, L291:33--36, 1985.

\bibitem{Shapiro:1983du}
S.~L. Shapiro and S.~A. Teukolsky.
\newblock {Black holes, white dwarfs, and neutron stars: The physics of compact
  objects}.
\newblock 1983.

\bibitem{Shlapentokh-Rothman:2013ysa}
Y.~Shlapentokh-Rothman.
\newblock {Exponentially growing finite energy solutions for the Klein-Gordon
  equation on sub-extremal Kerr spacetimes}.
\newblock {\em Commun.Math.Phys.}, 329:859--891, 2014.

\bibitem{Sperhake:2012me}
U.~Sperhake, E.~Berti, V.~Cardoso, and F.~Pretorius.
\newblock {Universality, maximum radiation and absorption in high-energy
  collisions of black holes with spin}.
\newblock {\em Phys.Rev.Lett.}, 111(4):041101, 2013.

\bibitem{Sperhake:2008ga}
U.~Sperhake, V.~Cardoso, F.~Pretorius, E.~Berti, and J.~A. Gonzalez.
\newblock {The high-energy collision of two black holes}.
\newblock {\em Phys. Rev. Lett.}, 101:161101, 2008.

\bibitem{Sperhake:2009jz}
U.~Sperhake et~al.
\newblock {Cross section, final spin and zoom-whirl behavior in high- energy
  black hole collisions}.
\newblock {\em Phys. Rev. Lett.}, 103:131102, 2009.

\bibitem{stewart}
J.~M. Stewart.
\newblock {Solutions of the wave equation on a Schwarzschild space-time with
  localized energy}.
\newblock {\em Proc. R. Soc. London}, A424:239--244, 1989.

\bibitem{Teukolsky:2014vca}
S.~A. Teukolsky.
\newblock {The Kerr Metric}.
\newblock 2014.
\newblock arXiv:1410.2130 [gr-qc].

\bibitem{Vishveshwara:1970cc}
C.~V. Vishveshwara.
\newblock {Stability of the schwarzschild metric}.
\newblock {\em Phys. Rev.}, D1:2870--2879, 1970.

\bibitem{Witek:2012tr}
H.~Witek, V.~Cardoso, A.~Ishibashi, and U.~Sperhake.
\newblock {Superradiant instabilities in astrophysical systems}.
\newblock {\em Phys.Rev.}, D87:043513, 2013.

\bibitem{Yoshino:2012kn}
H.~Yoshino and H.~Kodama.
\newblock {Bosenova collapse of axion cloud around a rotating black hole}.
\newblock {\em Prog.Theor.Phys.}, 128:153--190, 2012.

\bibitem{Yunes:2011aa}
N.~Yunes, P.~Pani, and V.~Cardoso.
\newblock {Gravitational Waves from Quasicircular Extreme Mass-Ratio Inspirals
  as Probes of Scalar-Tensor Theories}.
\newblock {\em Phys.Rev.}, D85:102003, 2012.

\bibitem{Zerilli:1971wd}
F.~J. Zerilli.
\newblock {Gravitational field of a particle falling in a schwarzschild
  geometry analyzed in tensor harmonics}.
\newblock {\em Phys. Rev.}, D2:2141--2160, 1970.

\bibitem{Zouros:1979iw}
T.~Zouros and D.~Eardley.
\newblock {Instabilities of Massive Scalar Perturbations of a Rotating Black
  Hole}.
\newblock {\em Annals Phys.}, 118:139--155, 1979.

\end{thebibliography}

\end{document}